\documentclass[
twocolumn,
tightenlines,
10pt,
longbibliography,
showpacs,
nofootinbib,
notitlepage,
superscriptaddress]{revtex4-2}

\usepackage{multibib}
\usepackage{amsmath, amssymb, amsthm, braket, enumitem, dsfont, relsize}
\usepackage{makecell, graphicx, subfiles, multirow, wrapfig}
\usepackage[font=scriptsize]{subcaption}
\usepackage[font=scriptsize, justification=raggedright]{caption}
\usepackage{subcaption}
\usepackage[page]{appendix}
\usepackage[colorlinks, allcolors=black]{hyperref}
\usepackage[capitalize, nameinlink]{cleveref}
\usepackage{xcolor, tabularx}
\usepackage{tikz}
\usetikzlibrary{quantikz}
\usetikzlibrary{shapes.geometric, arrows}
\usetikzlibrary{positioning} 
\setlist[itemize]{leftmargin=10pt}

\DeclareCaptionLabelFormat{bold}{\textbf{(#2)}}
\begin{document}

\title{Neural Error Mitigation of Near-Term Quantum Simulations}

\author{Elizabeth~R.~Bennewitz}
\email[]{E. R. B and F. H. have contributed equally to this work.}
\affiliation{1QB Information Technologies (1QBit), Waterloo, ON, Canada}
\affiliation{Department of Physics \& Astronomy, University of Waterloo, Waterloo, ON, Canada}
\affiliation{Perimeter Institute for Theoretical Physics, Waterloo, ON, Canada}

\author{Florian~Hopfmueller}
\email[]{E. R. B and F. H. have contributed equally to this work.}
\affiliation{1QB Information Technologies (1QBit), Waterloo, ON, Canada}
\affiliation{Department of Physics \& Astronomy, University of Waterloo, Waterloo, ON, Canada}
\affiliation{Perimeter Institute for Theoretical Physics, Waterloo, ON, Canada}

\author{Bohdan~Kulchytskyy}
\affiliation{1QB Information Technologies (1QBit), Waterloo, ON, Canada}

\author{Juan~Carrasquilla}
\affiliation{Vector Institute, MaRS Centre, Toronto, ON, Canada}
\affiliation{Department of Physics \& Astronomy, University of Waterloo, Waterloo, ON, Canada}

\author{Pooya~Ronagh}
\thanks{{\vskip-10pt}{\hskip-8pt}Corresponding author: \href{mailto:pooya.ronagh@1qbit.com}{pooya.ronagh@1qbit.com}\\}
\affiliation{Institute for Quantum Computing, University of Waterloo, Waterloo, ON, Canada}
\affiliation{Department of Physics \& Astronomy, University of Waterloo, Waterloo, ON, Canada}
\affiliation{Perimeter Institute for Theoretical Physics, Waterloo, ON, Canada}
\affiliation{1QBit, Vancouver, BC, Canada}

\date{\today}

\begin{abstract}
Near-term quantum computers provide a promising platform for finding ground states of quantum systems, which is an 
essential task in physics, chemistry, and materials science. Near-term approaches, however, are constrained by the effects of noise 
as well as the limited resources of near-term quantum hardware. We introduce \textit{neural error mitigation}, which uses 
neural networks to improve estimates of ground states and ground-state observables obtained using near-term quantum 
simulations. To demonstrate our method’s broad applicability, we employ neural error mitigation to find the ground states of 
the H$_\text{2}$ and LiH molecular Hamiltonians, as well as the lattice Schwinger model, prepared via the variational quantum eigensolver 
(VQE). Our results show that neural error mitigation improves numerical and experimental VQE computations to yield low 
energy errors, high fidelities, and accurate estimations of more-complex observables like order parameters and 
entanglement entropy, without requiring additional quantum resources. Furthermore, neural error mitigation is agnostic with respect 
to the quantum state preparation algorithm used, the quantum hardware it is implemented on, and the particular noise channel affecting 
the experiment, contributing to its versatility as a tool for quantum simulation.
\end{abstract}

\maketitle


\section{Introduction}
Since the early twentieth century, scientists have been developing comprehensive
theories that describe the behaviour of quantum mechanical systems. However, the
computational cost required to study these systems often exceeds the
capabilities of current scientific computing methods and hardware. Consequently,
computational infeasibility remains a roadblock for the practical application of
those theories to problems of scientific and technological importance.

The simulation of quantum systems on quantum computers, referred to in this
paper as quantum simulation, shows promise toward overcoming these roadblocks,
and has been a foundational driving force behind the conception and creation of
quantum computers~\cite{feynman1982simulating, bennett1973logical,
benioff1980computer, manin1980vychislimoe}. In particular, the quantum
simulation of ground and steady states of quantum many-body systems beyond the
capabilities of classical computers is expected to significantly impact nuclear
physics, particle physics, quantum gravity, condensed matter physics, quantum
chemistry, and materials science~\cite{preskill2018simulating, cao2019quantum,
mcardle2020quantum, bauer2020quantum}. The capabilities of current and near-term
quantum computers continue to be constrained by limitations, such as the number
of qubits and the effects of noise. Quantum error correction (QEC) techniques
can eliminate errors that result from noise, providing a path toward
fault-tolerant quantum computation. However, in practice, implementing QEC
imposes a large overhead in terms of both the required number of qubits and low
error rates, both of which remain beyond the capabilities of current and
near-term devices.

Before fault-tolerant quantum simulations~\cite{aspuru2005simulated} can be
realized, modern variational algorithms significantly alleviate the demand on
quantum hardware and exploit the capabilities of noisy intermediate-scale
quantum (NISQ) devices \cite{bharti2021noisy, cerezo2021variational}.
 A prominent example is the variational quantum
eigensolver (VQE)~\cite{peruzzo2014variational}, a hybrid quantum--classical
algorithm that iteratively approximates the lowest-energy eigenvalues of a
target Hamiltonian through the variational optimization of a family of
parameterized quantum circuits. This, and other variational algorithms, has
emerged as a leading strategy toward achieving a quantum advantage using
near-term devices and accelerating progress in multiple scientific and
technological fields~\cite{endo2021hybrid}.

The experimental implementation of variational quantum algorithms remains a
challenge for many scientific problems, as NISQ devices suffer from various
sources of noise and imperfection. To alleviate these issues, several methods
for quantum error mitigation (QEM) have been proposed and experimentally
validated that improve quantum computations in the absence of the quantum
resources required for QEC~\cite{roffeQuantumErrorCorrection2019}. For a review
of current QEM techniques, we refer the reader to Ref.~\cite{endo2021hybrid} and
the material cited therein. In general, these methods use specific information
about the noise channels that affect a quantum computation, the hardware
implementation, or the quantum algorithms themselves. Examples include the
implicit characterization of noise models and how they affect estimates of the
desired observables, specific knowledge of the state subspaces in which the
prepared quantum state ought to reside, and the characterization and mitigation
of the sources of noise on individual components of the quantum computation such
as single- and two-qubit gate errors, as well as state preparation and
measurement (SPAM) errors.

\begin{figure}
\centering
\includegraphics[width=\linewidth]{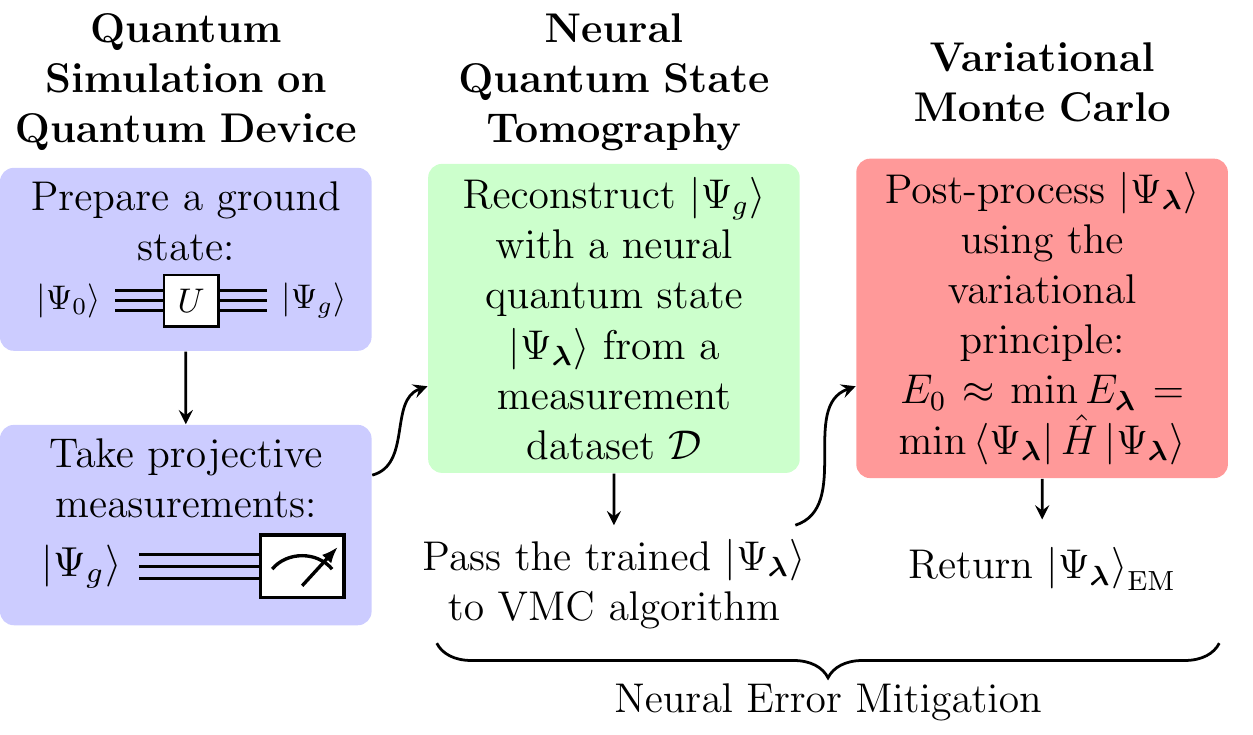}
\caption{
\textbf{Neural error mitigation procedure $|$} First, an approximate ground state $\ket{
\Psi_g}$ is prepared on a quantum computer from which simple
projective measurements are taken (left column). This measurement dataset,
$\mathcal{D}$, is then used to reconstruct the final state $\ket{\Psi_g}$ with a
neural quantum state $\ket{\Psi_{\vec \lambda}}$ using neural quantum state
tomography (middle column). Then, the neural network ansatz is post-processed
using variational Monte Carlo to mitigate errors in the ground-state
representation (right column).}
\label{fig:NEM flow chart}
\end{figure}

Machine learning techniques, which have recently been repurposed as tools for
tackling complex problems in quantum many-body physics and quantum information
processing~\cite{dunjkoMachineLearningArtificial2018,
carrasquillaMachineLearningQuantum2020a}, provide an alternative route to QEM.
Here we introduce a QEM strategy named \emph{neural error mitigation} (NEM),
which uses neural networks to mitigate errors in the approximate preparation of
the quantum ground state of a Hamiltonian.

The NEM algorithm, summarized in~\cref{fig:NEM flow chart}, is composed of two
steps. First, we perform neural quantum state tomography (NQST) to train a
neural quantum state (NQS) ansatz to represent the approximate ground state
prepared by a noisy quantum device, using experimentally accessible
measurements. Inspired by traditional quantum state tomography (QST), NQST is a
data-driven machine learning approach to QST that uses a limited number of
measurements to efficiently reconstruct complex quantum
states~\cite{torlai2018neural}. We then apply the variational Monte Carlo
algorithm (VMC) on the same neural quantum state ansatz (which we call the NEM
ansatz) to improve the representation of the unknown ground state. In the spirit
of VQE, VMC approximates the ground state of a Hamiltonian based on a classical
variational ansatz~\cite{becca2017quantum}, in this case a NQS ansatz.

In this paper, we use an autoregressive generative neural network as our NEM
ansatz. In particular, we use the Transformer~\cite{vaswani2017attention}
architecture, and show that this model performs well as a neural quantum state.
Due to its capability to model long-range temporal and spatial correlations,
this architecture has led to many state-of-the-art results in natural language
and image processing, and has the potential to model long-range quantum
correlations. We refer the reader to the Methods section and Supplementary
Information for a complete description of NQS, NQST, VMC, and the Transformer
neural network.

Neural error mitigation has several advantages compared to other error
mitigation techniques. Firstly, it has a low experimental overhead; it requires
only a set of simple experimentally feasible measurements to learn the
properties of the noisy quantum state prepared by VQE. Consequently, the
overhead of error mitigation in NEM is shifted from quantum
resources (i.e., performing additional quantum experiments and measurements) 
to classical computing resources for machine learning.
In particular, we note that the primary cost of NEM is in performing VMC until convergence.
 Another advantage of NEM is that it is agnostic with respect to the quantum
simulation algorithm, the device it is implemented on, and the particular noise
channel affecting the quantum simulation. As a result, it can also be combined
with other QEM techniques~\cite{cai2020multi, torlai2019integrating}, and can be
applied to either analog quantum simulation or digital quantum
circuits~\cite{song2019quantum, sun2020mitigating}.

Neural error mitigation also alleviates the low measurement precision that
arises when estimating quantum observables using near-term quantum devices. This
is particularly important in quantum simulations, where making accurate
estimations of quantum observables is essential for practical applications.
Neural error mitigation intrinsically resolves the low measurement precision at
each step of the algorithm. During the first step, NQST improves the variance of
observable estimates at the cost of introducing a small estimation
bias~\cite{torlai2020precise}. This bias, as well as the remaining variance, is
further reduced by training the NEM ansatz using VMC, which results in a
zero-variance expectation value for energy estimates once the ground state has
been reached~\cite{assaraf1999zero}.

\begin{figure*}
\centering
  \begin{subfigure}{0.32\textwidth}
    \centering
    \caption{}
    \includegraphics[width=.99\linewidth]
    {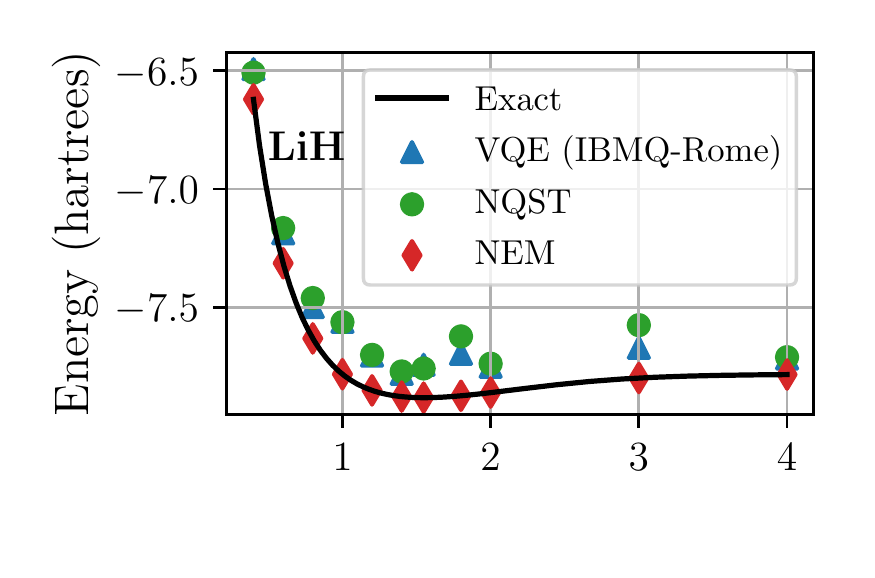}
    \label{fig:LiH Energy (ibmq_rome)}
  \end{subfigure}
  \begin{subfigure}{0.32\textwidth}
    \centering
    \caption{}
    \includegraphics[width=.99\linewidth]
    {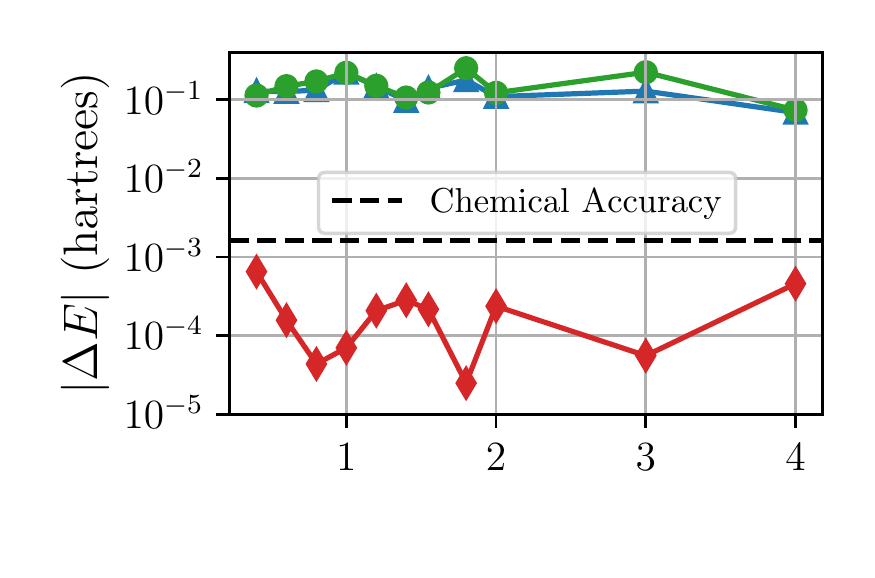}
    \label{fig:LiH Energy Error (ibmq_rome)}
  \end{subfigure}
  \begin{subfigure}{0.32\textwidth}
    \centering
    \caption{}
    \includegraphics[width=.99\linewidth]
    {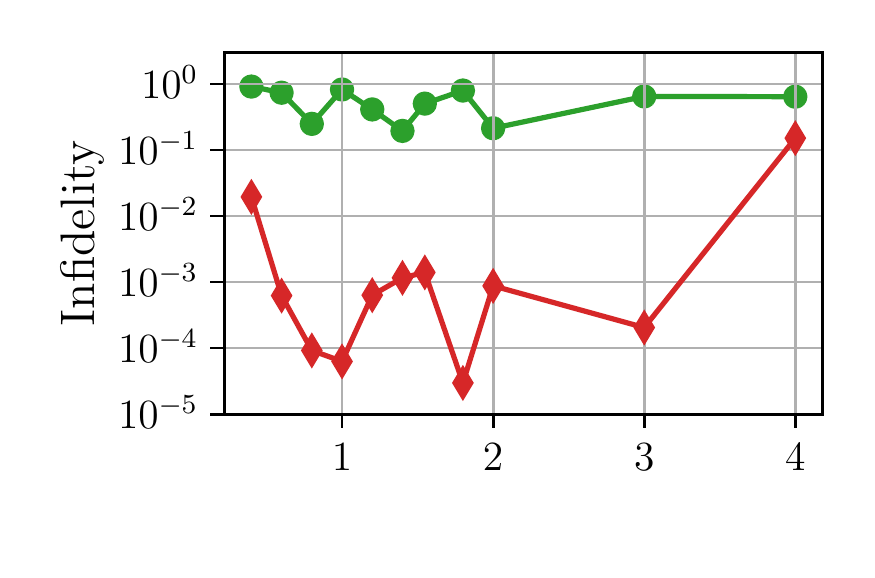}
    \label{fig:LiH Infidelity (ibmq_rome)}
  \end{subfigure}
  \\[-9ex]
  \begin{subfigure}{0.32\textwidth}
    \centering
    \caption{}
    \includegraphics[width=.99\linewidth]
    {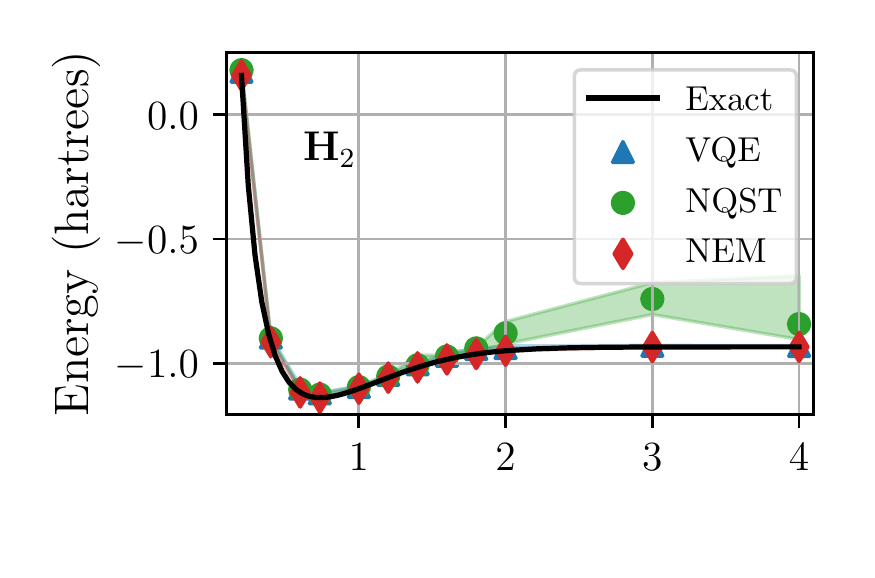}
    \label{fig:H2 Energy}
  \end{subfigure}
  \begin{subfigure}{0.32\textwidth}
    \centering
    \caption{}
    \includegraphics[width=.99\linewidth]
    {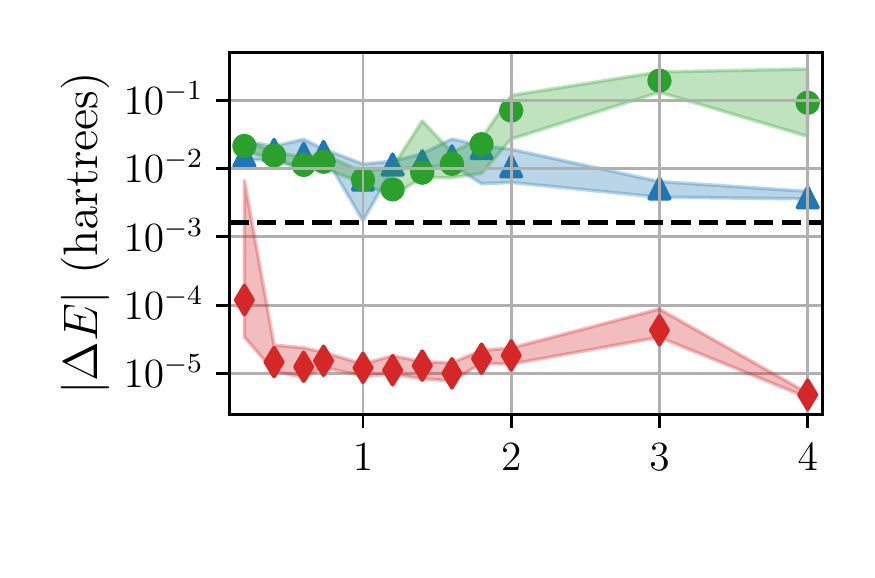}
    \label{fig:H2 Energy Error}
  \end{subfigure}
  \begin{subfigure}{0.32\textwidth}\quad
    \centering
    \caption{}
    \includegraphics[width=.99\linewidth]
    {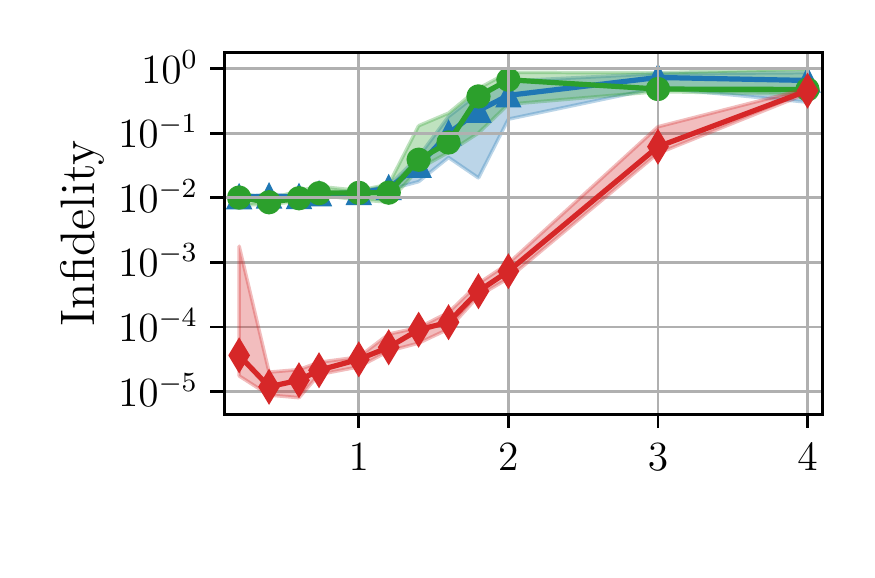}
    \label{fig:H2 Infidelity}
  \end{subfigure}
  \\[-9ex]
  \begin{subfigure}{0.32\textwidth}
    \centering
    \caption{}
    \includegraphics[width=.99\linewidth]
    {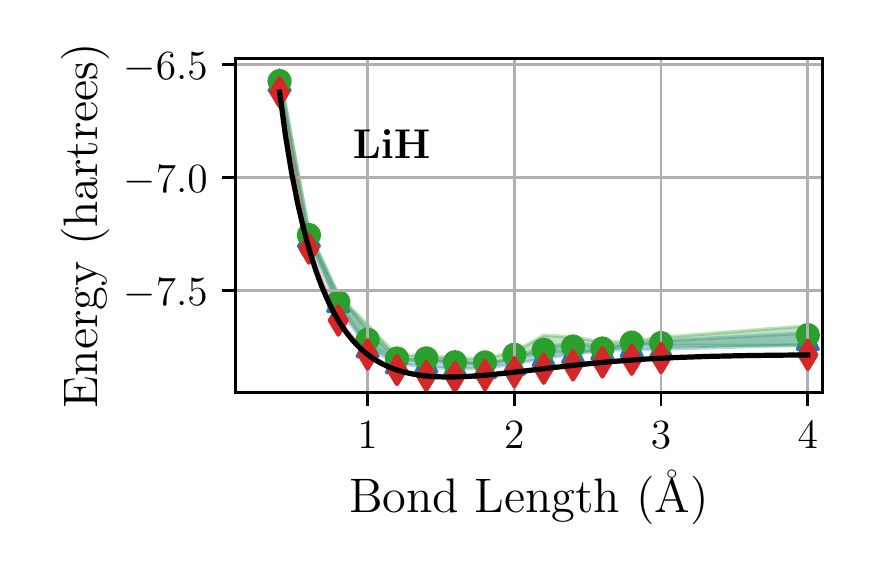}
    \label{fig:LiH Energy}
  \end{subfigure}
  \begin{subfigure}{0.32\textwidth}
    \centering
    \caption{}
    \includegraphics[width=.99\linewidth]
    {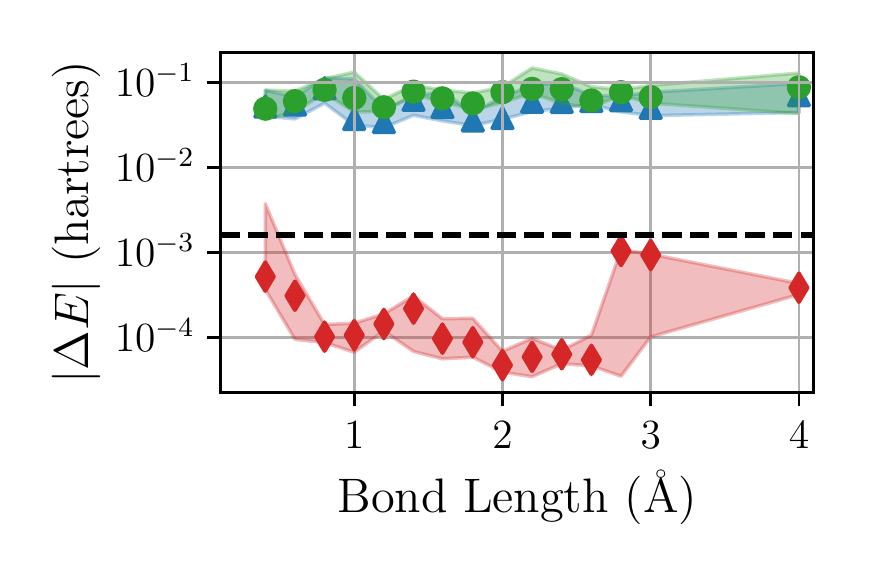}
    \label{fig:LiH Energy Error}
  \end{subfigure}
  \begin{subfigure}{0.32\textwidth}
    \centering
    \caption{}
    \includegraphics[width=.99\linewidth]
    {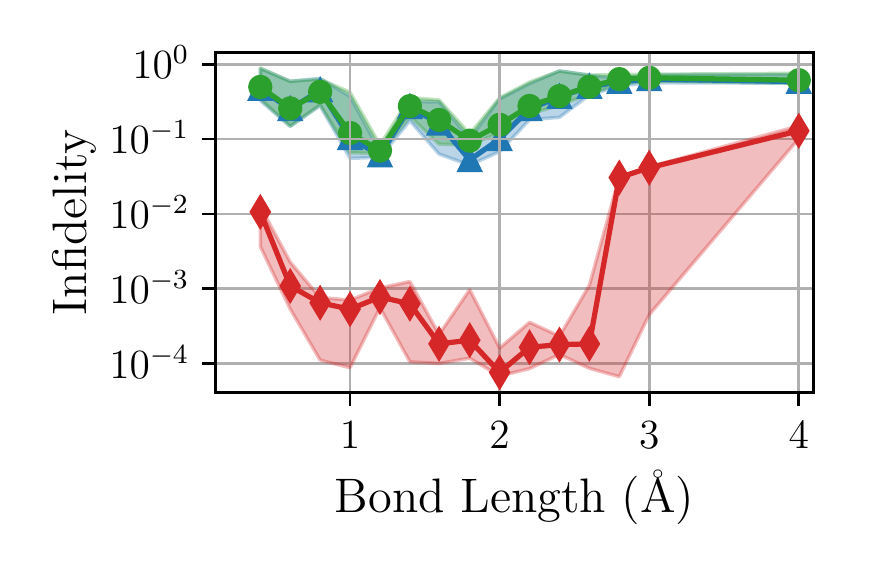}
    \label{fig:LiH Infidelity}
  \end{subfigure}
\\[-3ex]
\caption{
\textbf{Neural error mitigated experimental and numerical results for molecular
Hamiltonians $|$} Neural error mitigation results for energy, energy error, and
infidelity for the ground states of LiH and H$_2$ prepared using a
hardware-efficient variational quantum circuit. Each panel contains information
about the prepared quantum states (blue triangles), neural quantum states
trained using neural quantum state tomography (green circles), the final neural
error mitigated states (red diamonds), and, where appropriate, the exact results
(solid black line). The top row~\textbf{(a - c)} shows results for the LiH ground
state prepared experimentally using IBM's five-qubit device, IBMQ-Rome. The
middle row~\textbf{(d - f)} and bottom
row~\textbf{(g - i)} show the performance
of neural error mitigation for numerically prepared H$_2$ and LiH ground states,
respectively. Results are shown for the median performance over 10 noisy
numerical simulations per bond length, and the shaded region is the 
interquartile range. 
For our ten data points, the interquartile range includes the middle six and
excludes the best two and worst two in order to indicate the typical performance of the method. 
Error mitigated results extend both the
experimentally and numerically prepared VQE states to chemical accuracy and low
infidelities for all bond lengths of LiH and H$_2$. Chemical accuracy is shown at
$0.0016$ hartrees (dashed black line).}
\label{fig:quantum_chemistry_error_mitigation_performance}
\end{figure*}

\section{Results}
\subsection{Quantum Chemistry Results}
Accurately simulating a molecule's electron correlations is an integral step in
characterizing the chemical properties of the molecule. This problem, known as
the electronic structure problem, involves finding the ground-state wavefunction
and energy of many-body interacting fermionic molecular Hamiltonians. Achieving
an absolute energy error corresponding to chemical accuracy (1 kcal/mol
$\approx0.0016$ hartrees), the threshold for accurately estimating
room-temperature chemical reaction rates, is essential to applications in drug
discovery and materials science~\cite{mcardle2020quantum}.

We demonstrate the application of NEM to the estimation of molecular ground
states prepared using a VQE algorithm and show that our method improves the
results up to chemical accuracy or better for the H$_2$ and LiH molecules (see
\cref{fig:quantum_chemistry_error_mitigation_performance} for experimental and
numerical results). We map the H$_2$ and LiH molecular Hamiltonians 
computed in the STO-3G basis to qubit
Hamiltonians with $N = 2$ and $N = 4$ qubits,
respectively~\cite{kandala2017hardware}. The prepared quantum state is the
hardware-efficient variational quantum circuit composed of single-qubit Euler
rotation gates and two-qubit CNOT entangling gates native to superconducting
hardware~\cite{kandala2017hardware}. For both H$_2$ and LiH, we construct a
variational circuit with a single entangling layer, giving a variational circuit
with 10 and 20 parameters, respectively. More details can be found in the
Methods section.

We highlight the performance of NEM on the experimental preparation of the
ground states of LiH at different bond lengths using IBM's five-qubit chip,
IBMQ-Rome. We map the four-qubit LiH problem to the four linearly connected
qubits on IBMQ-Rome that have the lowest average single- and two-qubit gate
errors. During optimization, we perform 250 iterations of simultaneous
perturbation stochastic approximation (SPSA) optimization to obtain the final
prepared quantum state. Neural error mitigation improves the results of VQE to
chemical accuracy or better for all bond lengths, and achieves infidelities,
given by \mbox{$1-|\braket{\Psi | \Psi_0}|^2$}, of $10^{-3}$ for most bond
lengths (shown in top row
of~\cref{fig:quantum_chemistry_error_mitigation_performance}). On average, NEM
achieves an improvement of three orders of magnitude on energy estimation and
two orders of magnitude on infidelity. We provide further details about the
reconstruction quality in the Supplementary Information, including an analysis
of the reconstructed neural quantum state.

\begin{figure*}

\begin{tabular}{cc|c}
  \begin{subfigure}{.32\linewidth}
    \subcaption{}
    \includegraphics[width=\textwidth]{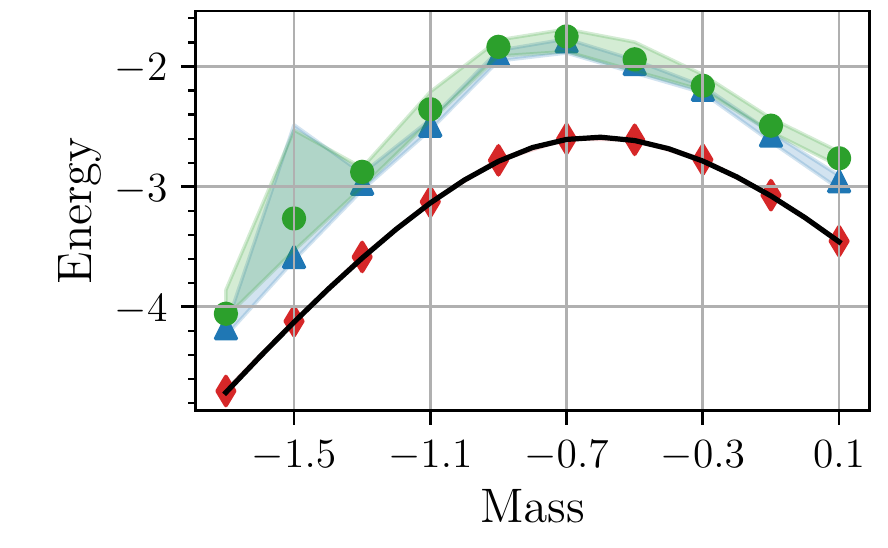}
    \label{fig:schwinger_errmit_performance_en}\\[-7ex]
    \subcaption{}
    \includegraphics[width=\textwidth]{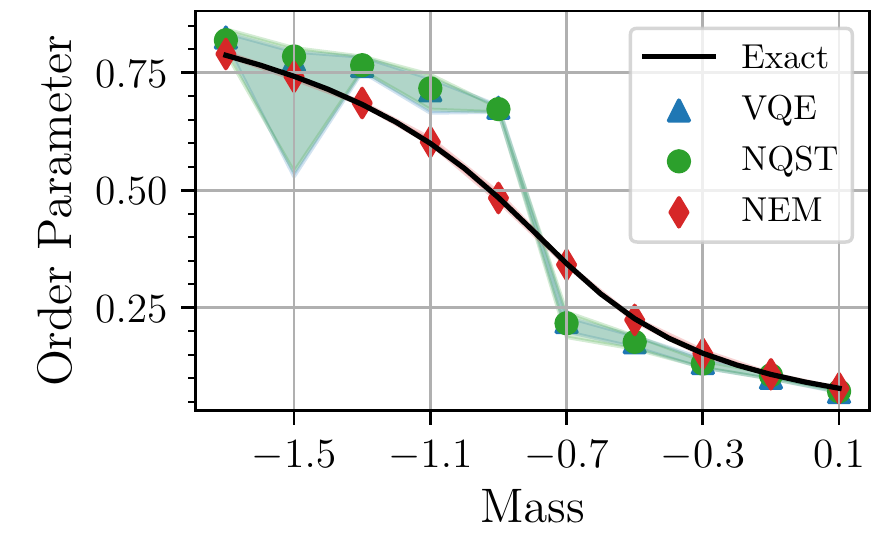}
    \label{fig:schwinger_errmit_performance_op}
  \end{subfigure} &

  \begin{subfigure}{.32\linewidth}
    \subcaption{}
    \includegraphics[width=\textwidth]{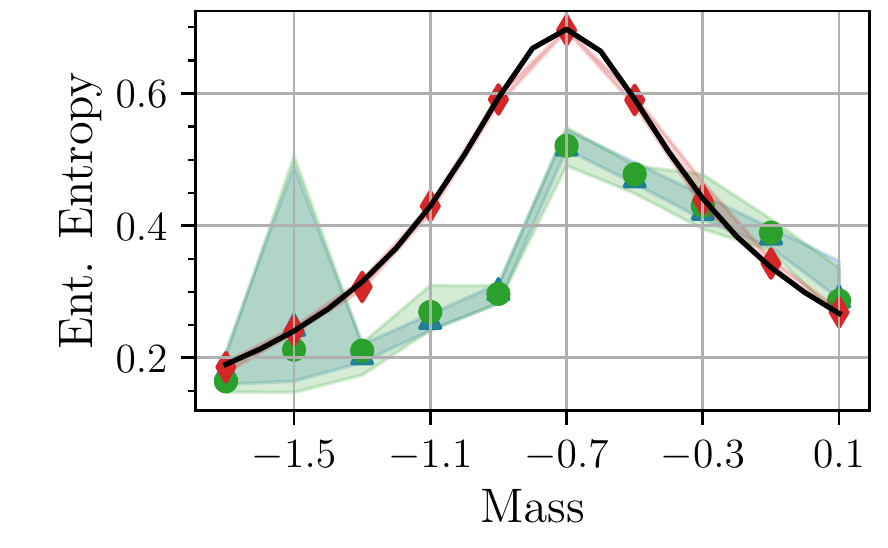}
    \label{fig:schwinger_errmit_performance_entent}\\[-7ex]
    \subcaption{}
    \includegraphics[width=\textwidth]{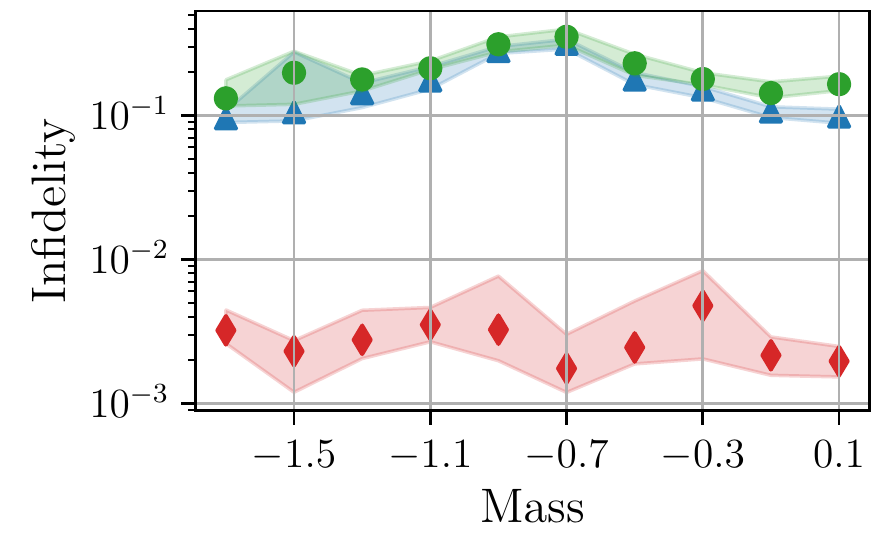}
    \label{fig:schwinger_errmit_performance_inf}
  \end{subfigure} &

  \begin{subfigure}{.32\linewidth}
    \subcaption{}
    \includegraphics[width=\textwidth]{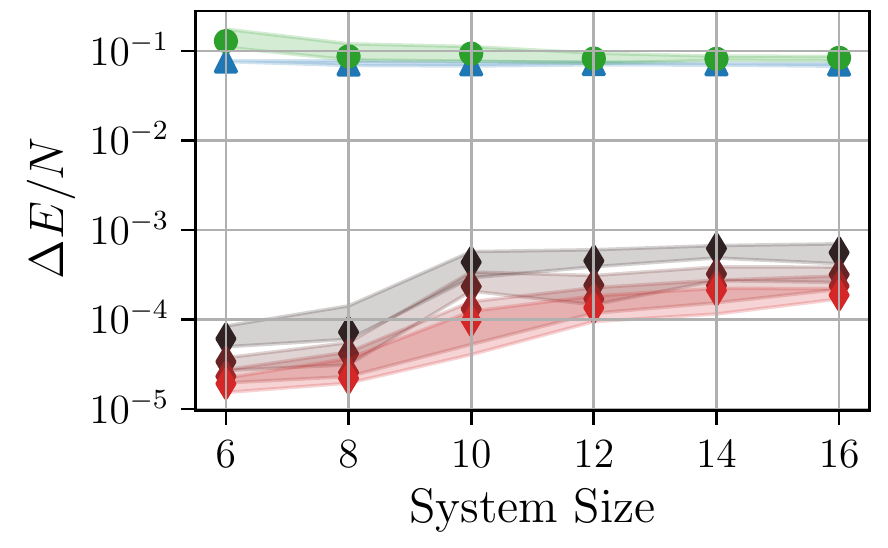}
    \label{subfig:schwinger_scaling_deltae}\\[-7ex]
    \subcaption{}
    \includegraphics[width=\textwidth]{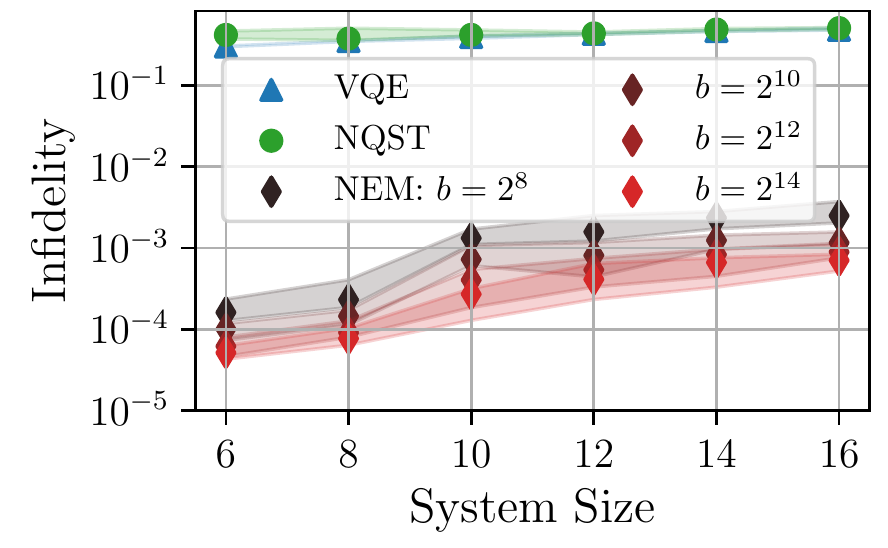}
    \label{subfig:schwinger_scaling_infid}
  \end{subfigure}
\end{tabular}
\\[-2ex]
\caption{\textbf{Performance of neural error mitigation applied to ground states
of the lattice Schwinger model $|$} \textbf{\textit{Left}}: Estimates
for~\textbf{(a)} the ground state
energy,~\textbf{(b)} order
parameter,~\textbf{(c)} entanglement entropy
between the first three and remaining five sites,
and~\textbf{(d)} infidelity to the exact ground
state, for the $N=8$ site model. Each panel contains results for the quantum
states prepared using VQE simulated with a depolarizing noise channel (blue
triangles), neural quantum states trained using neural quantum state tomography
(green circles), the final neural error mitigated neural quantum states (red
diamonds), and, where applicable, exact results (solid black lines). While the
qualitative behaviour of the entanglement entropy and the order parameter across
the phase transition are not modelled well by VQE or NQST, applying NEM
consistently improves the estimates of all observables to low errors and low
infidelities. \textbf{ \textit{Right}}: Results of the scaling study of NEM at
the phase transition ($m=-0.7$) are shown
for~\textbf{(e)} the energy error per site
and~\textbf{(f)} infidelity, as a function of system
size. The performance of VQE without noise (blue triangles) shows an
approximately constant energy error per site with infidelities that become
slightly worse as system size increases. Across all sizes, applying NEM (red
diamonds) improves VQE performance by two to four orders of magnitude, even when
using a small VMC batch size of $b=2^8$, which is the number of samples used to
estimate the energy's gradient in one iteration of VMC. In all panels (left and
right), median values over 10 runs are shown, and the shaded region is the interquartile range.}
\label{fig:schwinger_errmit_performance}
\end{figure*}

Additionally, we illustrate the results of applying NEM on the ground states of
H$_2$ and LiH prepared using classically simulated VQE with a depolarizing noise
channel (shown in bottom two rows
of~\cref{fig:quantum_chemistry_error_mitigation_performance}). We simulate VQE
with a single-qubit depolarizing error probability of $0.001$ and two-qubit
depolarizing error probability of $0.01$. At each bond length, we generate 10
VQE simulations and report the NEM results. Notably, the median performance of
NEM improves the ground-state estimation of H$_2$ and LiH to chemical accuracy and
low infidelities for all bond lengths. The increased infidelity for bond lengths
larger than $2$\r{A}, as shown in the infidelity plots
of~\cref{fig:quantum_chemistry_error_mitigation_performance}, can be explained
by the decreasing energy gap between the ground state and the first excited
state. When the energy gap is small, it becomes more difficult for methods that
optimize the energy, like VQE and VMC, to isolate the ground-state
representation.


\subsection{Lattice Schwinger Results}
We next apply our method to the ground state of the lattice Schwinger model,
which is a prototypical abelian lattice gauge theory, and a toy model for
quantum electrodynamics in one spatial dimension. Multiple experiments have been
proposed that use quantum devices to explore the properties of this model
\cite{kokail2019self,klco2018quantum, martinez2016real,
borzenkova2020variational}. In this paper, we consider the experiment where a
trapped-ion analog quantum simulator is used to variationally prepare the ground
state of the lattice Schwinger model using alternating entangling operations,
$e^{i t H_E}$, which act on all qubits simultaneously, and single qubit
rotations~\cite{kokail2019self}.

After using a Jordan--Wigner transformation to map the fermionic degrees of
freedom of the theory to qubits, the lattice Schwinger Hamiltonian takes the
following~\cite{kokail2019self} form:
\begin{align} \label{eq:schwinger_h}
  \hat H ={}& \frac{w}{2} \sum_{j=1}^{N-1} \big(\hat X_j \hat X_{j+1}
  + \hat Y_j \hat Y_{j+1} \big)
  + \frac m 2 \sum_{j=1}^N(-1)^j\hat Z_j \nonumber\\
  & + \bar g \sum_{j=1}^N \hat L_j^2.
\end{align}
The first term describes the creation and annihilation of electron--positron
pairs and contains an overall energy scale, $w$. The second term contains the
bare electron mass $m$, and the third term contains $\bar g$, which is the
coupling strength to the electric field $\hat L_j$. Solving for the electric
field in one spatial dimension gives
\mbox{$\hat L_j= \epsilon_0
- \frac{1}2 \sum_{\ell=1}^j\big(\hat Z_\ell + (-1)^\ell \hat I \big)$},
where $\epsilon_0$ is an integration constant. Given that the quantum fields at
one spatial lattice point are encoded into a pair of qubits, the total number of
sites $N$ must be even. We set $w=1, \bar g = 1$, and $\epsilon_0=0$ such that
the only remaining parameter is the mass $m$. The ground state of the system for
$m \rightarrow +\infty$ describes a vacuum with no electron--positron pairs and
for $m \rightarrow -\infty$ describes a large number of electron--positron
pairs. In the thermodynamic limit, the model exhibits a second-order phase
transition at $m \approx -0.7$, which can be detected using the order parameter
$\braket{\mathcal{O}}=
\frac{1}{2N(N-1)} \sum_{i, j>i}
\braket{(1+(-1)^i \hat Z_i)(1 + (-1)^j \hat Z_j)}$.
The model possesses discrete symmetries, which inform the choice of a
variational quantum circuit with a manageable number of parameters, but, in order
to demonstrate the general applicability of NEM, we
do not enforce these symmetries on the neural quantum state.

We demonstrate the performance of neural error mitigation by applying it to the
approximate ground state of the lattice Schwinger model obtained by numerically
simulating a VQE algorithm for $N=8$ sites, with single-qubit depolarizing noise
with probability $\lambda = 0.001$ applied after each rotation and entangling
operation. As shown in~\cref{fig:schwinger_errmit_performance}a
through~\cref{fig:schwinger_errmit_performance}d, the simple VQE scheme we
employ exhibits median infidelities between $0.10$ and $0.31$, with worse
performance closer to the phase transition around $m= -0.7$. While the
qualitative behaviour of the ground-state energy as a function of the mass is
modelled approximately by VQE, the qualitative behaviour of other physical
properties is not reproduced well, limiting the utility of our VQE results for studying the
phase transition. This includes the order parameter, and the Renyi entanglement
entropy $S_2$ of a partition of the system, which is a broadly used,
experimentally accessible quantity~\cite{islam2015measuring, brydges2019proving}  
expressing the amount of correlation present in the quantum state.

The properties of the NEM state show a substantial improvement over VQE. The NEM
state reaches absolute energy errors on the order of $10^{-2}$, and infidelities
approaching $10^{-3}$. Importantly, after applying NEM, the physical properties
estimated by the state accurately follow their exact values. The ability to
obtain precise estimations of these physical properties can be explained by the
accurate representation of the ground-state wavefunction captured by the NEM
neural quantum state. Further details about the reconstruction quality of each
component are covered in the Supplementary Information, including a thorough
analysis of the NEM neural quantum state.

To gather evidence that the performance of NEM scales well to larger near-term
experiments on quantum devices, we study the behaviour of NEM as a function of
system size for the lattice Schwinger model. For computational efficiency, the
scaling study uses a modified VQE implementation without noise
(see~\cref{fig:schwinger_errmit_performance}e
and~\cref{fig:schwinger_errmit_performance}f) as compared to the simulated
trapped-ion experiment (see~\cref{fig:schwinger_errmit_performance}a
through~\cref{fig:schwinger_errmit_performance}d). For more details on the
modified circuit, we refer the reader to the Methods section. The VQE algorithm
is simulated on a classical computer for system sizes up to $N= 16$, and NEM is
applied to the resulting states. The results
in~\cref{fig:schwinger_errmit_performance}e
and~\cref{fig:schwinger_errmit_performance}f show that NEM improves upon the VQE
results by two to four orders of magnitude, even for large system sizes, using
modest classical resources for the estimation of energy gradients in VMC.

\section{Discussion}
The error mitigation strategy developed here demonstrates significant improvements 
to the estimations of ground states and ground-state observables obtained from two 
example classes of near-term quantum simulations, independent of the quantum 
device and noise channels. Additionally, we
show that NEM exhibits the potential to scale well for
such quantum experiments. Given its low quantum overhead, NEM can be a powerful
asset for the error mitigation of near-term quantum simulations.

An advantage of using techniques based on neural quantum states for the task of
quantum error mitigation is the ability to approximate complex wavefunctions
from simple experimental measurements. In the process of improving the energy
estimation performed by VQE, NEM reconstructs and improves the ground-state
wavefunction itself as a neural quantum state. The accurate final representation
of the ground-state wavefunction is the reason why NEM is able to accurately
reconstruct and improve estimations of complex observables like energy, order
parameters, and entanglement entropy without imposing additional quantum
resources.

By combining VQE, which uses a parametric quantum circuit as an ansatz, and NQST
and VMC, which use neural networks as an ansatz, NEM brings together two
families of parametric quantum states and three optimization problems over their
loss landscapes~\cite{huembeli2021characterizing, park2020geometry,
bukov2020learning}. Our work raises the question as to the nature of the
relationships between these families of states, their loss landscapes, and
quantum advantage. Examining these relationships offers a new way to investigate
the potential of NISQ algorithms in seeking a quantum advantage. This may lead
to a better delineation between classically tractable simulations of quantum
systems and those that require quantum resources.


\section{Methods}
\label{sec:methods}

\subsection{Neural Quantum State}
\label{sec:methods:nnqs}

Our neural quantum state is based on the
Transformer~\cite{vaswani2017attention}, an architecture originally developed to
process sequences that have temporal and spatial correlations, such as written
languages. Compared to previous architectures for sequence models such as the
long short-term memory (LSTM) neural network~\cite{hochreiter1997long}, the
Transformer excels at modelling long-range correlations, and has thus become
very popular in machine learning. Within the quantum many-body machine learning
community, there has been a lot of work using autoregressive neural networks as
neural quantum states~\cite{hibat2020recurrent, sharir2020deep}. Recently, the
Transformer has been adapted as an autoregressive generative neural quantum
state~\cite{carrasquilla2019probabilistic}.

We represent the quantum state $|\psi \rangle$ with a Transformer neural network
that takes as input a bitstring $s = (s_1,\ldots,s_N) \in \{0,1\}^N$, describing a
computational basis state $|s\rangle$, where $N$ is the number of qubits. The neural network 
outputs two numbers
($p_{\vec \lambda}(s), \varphi_{\vec \lambda}(s)$) parameterized by the neural
network weights ${\vec \lambda}$, which form the complex amplitude
$\braket{ s | \psi_{\vec \lambda} }$ given by
\begin{equation} \label{NNQST:Mapping output to wavefunction}
  \langle s | \psi_{\vec \lambda} \rangle
  = \sqrt{p_{\vec \lambda}(s)} e^{i \varphi_{\vec \lambda}(s)}.
\end{equation}
Here, $p_{\vec \lambda}(s)$ is a normalized probability distribution, which
automatically normalizes the quantum state. The autoregressive property of the
model allows for efficient sampling from the Born distribution of
$\ket{\psi_{\vec \lambda}}$ in the computational basis. More details may be
found in the Supplementary Information.

The exact ground state amplitudes of both the quantum chemistry models and the
lattice Schwinger model are real, that is, $\varphi(s) \in \{0, \pi\}$, and the
signs of the lattice Schwinger model ground-state amplitudes follow a simple
sign rule. However, to show the general applicability of our method, we do not
enforce any of these conditions in our neural quantum states.

\subsection{Neural Quantum State Tomography}
\label{sec:methods:nqst}

In neural quantum state tomography \cite{torlai2018neural}, a neural network is
trained to represent the state of a quantum device using samples from that state
in various Pauli bases (i.e., after performing various post-rotations). Neural
quantum state tomography proceeds by iteratively adjusting the NQS parameters to
maximize the likelihood that NQS assigns to the samples.

A sample $s \in \{0, 1\}^N$ in a Pauli basis $B = P_1 P_2 \cdots P_N$, with $P_i
\in \{X, Y, Z\}$, is the unique simultaneous eigenstate of the single qubit
Pauli operators $P_i$, with eigenvalues determined by the entries of $s$. We
denote such a state by $\ket{s,B}$. The likelihood of the sample $(s, B)$
according to the neural quantum state $\ket{\psi_{\vec \lambda}}$ is given by

\begin{align}
  p_{\vec \lambda}(s, B)
  &= \big| \braket{ s, B | \psi_{\vec \lambda} } \big|^2 \nonumber \\
  &= \Big| \sum_{\substack{ t \in \{0,1\}^N \\ \langle s, B| t \rangle \neq 0}}
  \Braket{ s, B| t } ~ \Braket{ t | \psi_{\vec \lambda} } ~ \Big|^2.
\end{align}
Here, we sum over the computational basis states $\ket{t}$ that have a non-zero
overlap with the given sample $\ket{s,B}$. For a single sample $\ket{s,B}$, the
number of these states $|t\rangle$ is $2^K$, where $K$ is the number of
positions $i$ where $P_i \neq Z$. Therefore, the computational cost of a single
iteration of tomography training is proportional to $2^K$. To constrain this
computational cost, we use projective measurements in \emph{almost-diagonal}
Pauli bases (i.e., Pauli bases $B$ with low numbers of $X$ or $Y$ terms).

In order to learn the quantum state from a set of measurements $\mathcal D$, the
objective function minimized during NQST is an approximation of the cross
entropy averaged over the set of bases $\mathcal{B}$ from which samples were
drawn~\cite{torlai2020precise}, and is given by

\begin{align}
  L_{\vec \lambda} =
  -\frac{1}{|\mathcal{B}|} \sum_{B \in \mathcal B} \sum_{s \in \{0, 1\}^N}
  p_{\text{VQE}}(s, B) \ln p_{\vec \lambda}(s, B).
\end{align}
Here, $p_{\text{VQE}}(s, B)$ is the exact, unknown likelihood of measuring
$\ket{s, B}$ from the VQE state. The cross entropy achieves its minimum in
$\vec\lambda$ if $p_{\text{VQE}}(s, B) = p_{\vec\lambda}(s, B)$. As commonly
done in unsupervised learning, the cross entropy is approximated using the set
$\mathcal{D}$ of the measured samples $\ket{s,B}$, which is further partitioned
into training and validation subsets $\mathcal{D}_{T, V}$. The loss function
used in training is

\begin{align} \label{NQST Loss}
  L_{\vec \lambda}
  \approx -\frac{1}{|\mathcal{D}_T|}
  \sum_{\ket{s, B} \in \mathcal D_T} \ln p_{\vec \lambda}(s, B).
\end{align}
The training is performed using stochastic gradient descent (SGD) with the Adam
\cite{kingma2014adam} optimizer.

\subsection{Variational Monte Carlo and Regularization}
\label{sec:methods:vmc}

Variational Monte Carlo is a method that adjusts the parameters of a (classical)
variational wavefunction ansatz in order to approximate the ground state of a
given Hamiltonian. The method usually proceeds by gradient-based optimization of
the energy, where the energy and its partial derivatives with respect to the
ansatz parameters are estimated using Monte Carlo samples drawn from the
classical variational wavefunction. As detailed in the Supplementary
Information, the autoregressive property of our neural network wavefunction
allows for efficient sampling of the learned probability distribution. This
leads to more efficient VMC training compared to models such as restricted
Boltzmann machines (RBM)~\cite{carleo2017solving, choo2020fermionic}, where
samples have to be obtained using Markov chain Monte Carlo.

Many implementations of VMC use second-order methods involving the Hessian of
the energy~\cite{otis2019complementary}, or other update rules such as
stochastic reconfiguration~\cite{carleo2017solving}, to update the parameters.
These methods tend to be computationally expensive for models with large numbers
of parameters. Instead, we use SGD via the Adam optimizer, leading to an
update-step cost that is linear in the number of parameters, and hence scales
more favourably for larger models.

We find it necessary to add a regularization term to the VMC objective in the
early stages of VMC optimization. Without it, the magnitudes of the amplitudes
of some computational basis states decrease to almost zero over the course of
training, even for computational basis states which have a non-zero overlap with
the true ground state. It has been previously noted~\cite{choo2020fermionic}
that the VMC algorithm has difficulty finding the ground state of molecular
Hamiltonians because they have sharply peaked amplitudes in a sparse subset of
the computational basis states. The regularization term is designed to increase
very small amplitudes. It maximizes the $L_1$ norm of the state, thereby
discouraging sparsity. This is in contrast to the common usage of $L_1$
regularization in machine learning and optimization where the $L_1$ norm is
minimized to encourage sparsity in sparse optimization tasks. We expect this technique to be
useful for systems where the ground state has a large overlap with one or a few
computational basis states. 
For example, ground states of electronic structure problems have a large overlap
with the Hartree--Fock state and the lattice Schwinger ground states have a large overlap with 
the ground states for $m \rightarrow \pm \infty$.
This regularization technique allows the NQS to capture the subdominant amplitudes,
rather than collapsing onto the dominant computational basis state early in the training process. The $L_1$ norm can
be estimated in a tractable manner because we use an autoregressive, generative
neural network as our neural quantum state, which allows for the exact sampling
of the learned probability distribution and is automatically normalized. More
details on the regularization term and the VMC algorithm can be found in the
Supplementary Information.

\begin{figure}[h]

  \begin{subfigure}{\linewidth}
    \caption{}
    \includegraphics[width=\linewidth]{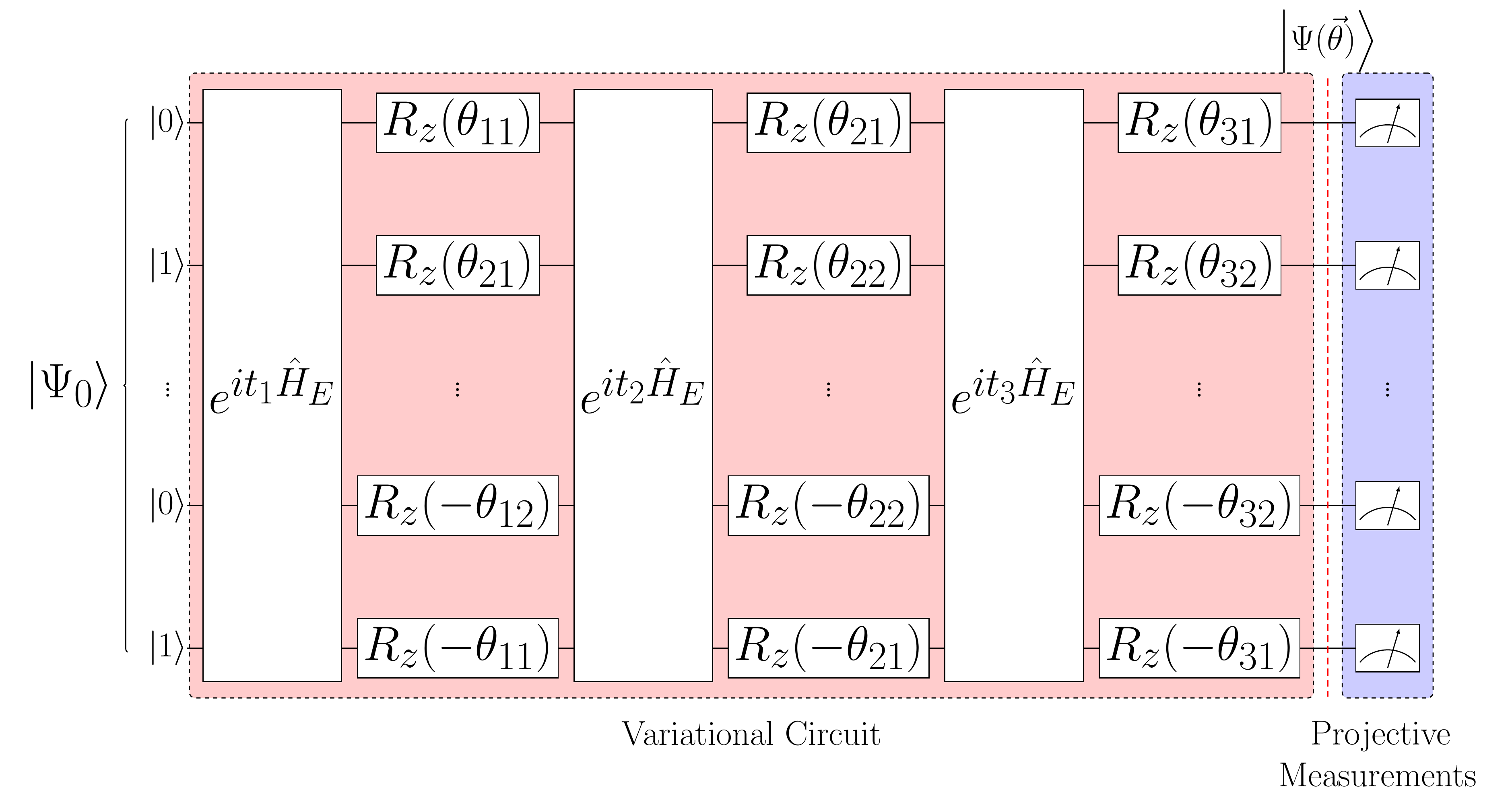}
    \label{fig:schwinger_vqe_circuit}
  \end{subfigure}

  \begin{subfigure}{\linewidth}
    \caption{}
    \includegraphics[width=.6\linewidth]{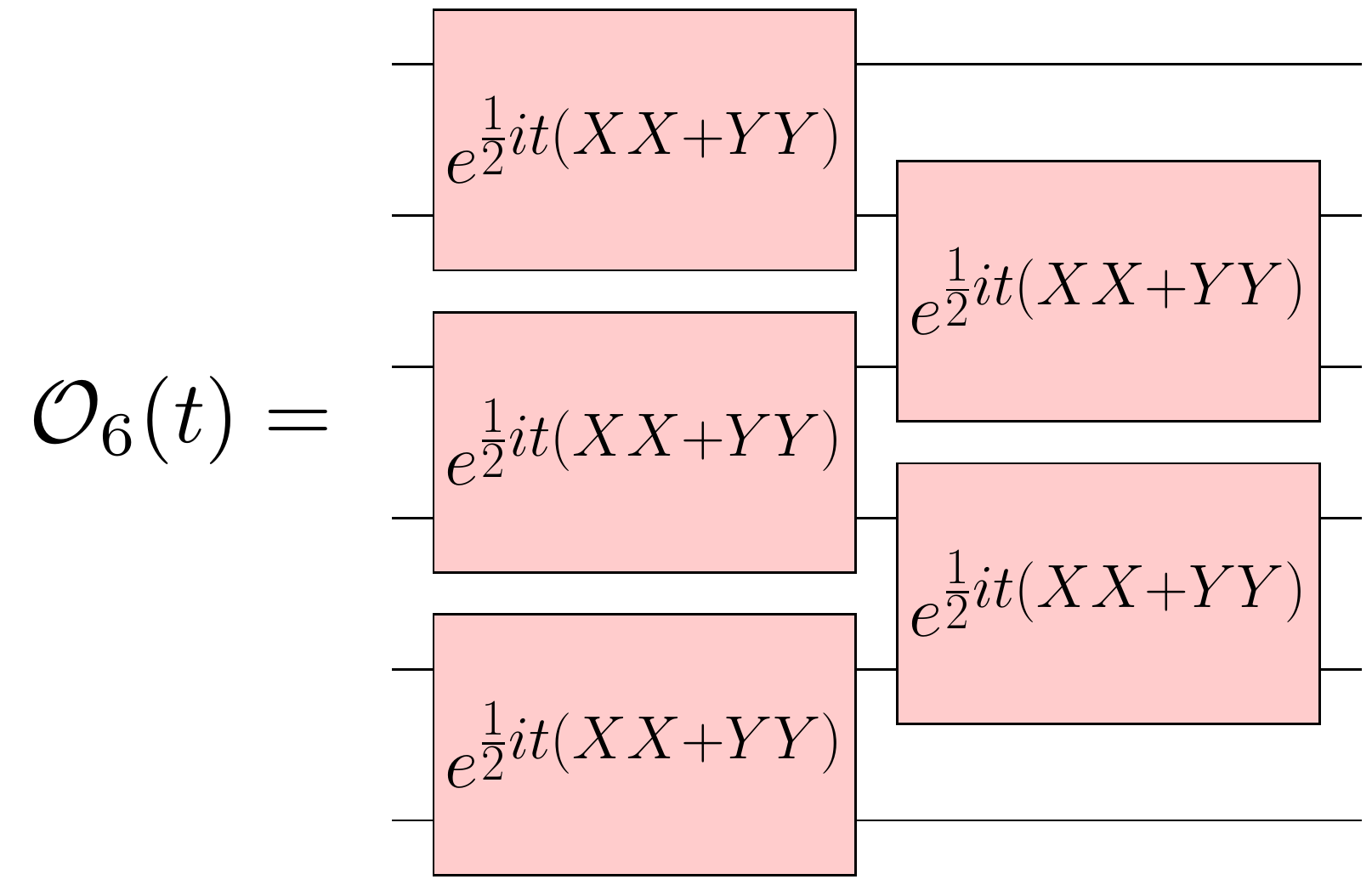}
    \label{fig:schwinger_scaling_entangling_gate}
  \end{subfigure}

\caption{
\textbf{VQE ansatz circuits for the lattice Schwinger model $|$}
\subref{fig:schwinger_vqe_circuit} Variational quantum circuit used to prepare
the approximate ground state of the lattice Schwinger model, using VQE simulated
classically. The input state $|\Psi_0 \rangle$ is $|01 \cdots 01\rangle$ ($|10
\cdots 10\rangle$) for $m \ge -0.7$ ($m < -0.7$).
\subref{fig:schwinger_scaling_entangling_gate} For the results shown
in~\cref{subfig:schwinger_scaling_deltae}
and~\cref{subfig:schwinger_scaling_infid}, the entangling layers
of~\subref{fig:schwinger_vqe_circuit} are replaced with $\mathcal{O}_N$ for $N$
sites. The layers of single-qubit gates are left unchanged.}
\end{figure}

\subsection{The VQE Implementation for Quantum Chemistry}
\label{sec:methods:quantum-chemistry-vqe-details}

We use the variational quantum circuit in \cref{sec:qc-variational-circuit} for
the electronic structure problem in quantum chemistry. This ansatz was designed
for the hardware capabilities of current superconducting quantum
processors~\cite{kandala2017hardware}. We use this circuit for both our
numerical simulations with a depolarizing noise channel, as well as to perform
experiments on a five-qubit superconducting quantum
processor~(\cref{fig:quantum_chemistry_error_mitigation_performance}).

The H$_2$ and LiH molecular Hamiltonians are mapped to qubit Hamiltonians with $N=
2$ and $N= 4$ qubits, respectively~\cite{kandala2017hardware}. In particular, we
map the second-quantized fermionic Hamiltonian for H$_2$ to its qubit Hamiltonian
using the Bravyi--Kitaev transformation~\cite{bravyi2002fermionic} while the LiH
Hamiltonian is transformed using the parity
transformation~\cite{seeley2012bravyi, bravyi2002fermionic}. In each case, two
qubits associated with the spin-parity symmetries of the model are removed to
obtain final qubit Hamiltonians~\cite{bravyi2017tapering}.

The hardware-efficient variational quantum circuit is composed of single-qubit
rotations and two-qubit entangling gates native to superconducting hardware. The
variational circuit,

\begin{align} \label{sec:qc-variational-circuit}
  \ket{\psi(\vec \theta)}
  = \prod_{l=1}^{d} \left( \prod_{q=1}^{N} \left[ U^{q,l}_{\rm EUR}
  (\vec \theta) \right] \times U_{\rm ENT} \right) \nonumber \\
  \times \prod_{q=1}^{N} \left[ U^{q,0}(\vec \theta) \right] \ket{00 \cdots 0},
\end{align}
for $N$ qubits consists of $d$ CNOT entangling layers alternating with $N(d+1)$
single-qubit Euler rotations, given by $U(\vec \theta) = R_Z(\theta_1)
R_X(\theta_2) R_Z(\theta_3)$. In the first rotation layer, $U^{q,0}(\vec \theta)
$, the first set of $Z$ rotations is not implemented, reducing the number of
circuit parameters. Within each entangling layer, we apply CNOT gates on pairs
of linearly connected qubits. The variational circuit has $p = N(3d+2)$
independent parameters. For both H$_2$ and LiH, we construct a variational circuit
with $d=1$ entangling layers giving $10$ and $20$ parameters, respectively.

The variational circuit is then optimized using Qiskit's implementation of
simultaneous perturbation stochastic approximation
(SPSA)~\cite{abraham2019qiskit} for 250 iterations to obtain an estimation for
the ground-state energy of H$_2$ and LiH. Each SPSA iteration requires two energy
evaluations. In order to reduce the sampling overhead during the energy
estimations, Pauli terms in each Hamiltonian are grouped according to their
common tensor product basis~\cite{kandala2017hardware}, requiring only two and
25 circuits with unique post-rotations for H$_2$ and LiH, respectively, to
estimate the energy.

In order to perform NQST, we collect almost-diagonal measurement samples from
the final noisy state prepared by the variational procedure. In this case, the
nearly diagonal samples are taken in the following bases: in the all-$Z$ basis,
in the $N$ bases with one $X$ (and $Z$ elsewhere), and in the $(N(N-1)/2)$ bases
with two $X$s (and $Z$ elsewhere).

\subsection{The VQE Implementation for the Lattice Schwinger Model}
\label{sec:methods:schwinger-vqe-details}

Our variational quantum circuit for the lattice Schwinger model closely follows
the variational circuit implementation on a trapped-ion analog quantum
simulator~\cite{kokail2019self} that approximately preserves the symmetries of
the lattice Schwinger model.

The quantum state is first prepared in $|01 \cdots 01\rangle$ for $m \ge 0.7$
and $|10 \cdots 10\rangle$ for $m < -0.7$, coinciding with the ground states of
the Schwinger Hamiltonian \cref{eq:schwinger_h} for $m \rightarrow \pm \infty$.
On this initial state, three alternating layers of evolution with an entangling
Hamiltonian, followed by $Z$ rotations on each qubit, are applied. The
entangling Hamiltonian contains long-range $XX$ couplings and a uniform
effective magnetic field, and is given by
\begin{equation}
  \hat H_E= J \sum_{j=1}^{N-1} \sum_{k=j+1}^N
  \frac{1}{|j-k|^\alpha}\hat X_j \hat X_k + B \sum_{j=1}^N \hat Z_j.
\end{equation}
We choose $\alpha = 1$, $J=1$, and $B=10$ to approximate the trapped-ion
experimental setup~\cite{kokail2019self}. Evolution with this Hamiltonian
preserves the symmetries of the lattice Schwinger model to first order terms in
$J/B$.

Only half of the parameters in each single-qubit rotation layer are independent,
as required by the symmetries, giving
$\varphi_j = - \varphi_{N+1-j}$ for $j \in \{N/2+1, \ldots, N\}.$
In total, the variational circuit has $15$ independent parameters for $N=8$
lattice sites, which are initialized to zero at the start of each optimization.
As a simple noise model, after each entangling layer and each single-qubit
rotation layer, a depolarizing channel with $\lambda = 0.001$ is applied to each
qubit (Extended Data Fig. 1a).

To optimize the variational parameters, the energy is estimated by taking
samples in each of the three bases $Z^{\otimes N}$, $X^{\otimes N}$, and
$Y^{\otimes N}$. The SPSA hyperparameter values are chosen by inspecting the
variance and approximate gradient at the beginning of the
optimization~\cite{spall1998implementation}. The exact values are listed in the
Supplementary Information.

As input to NQST, almost-diagonal samples are taken in each of the following
$2N-1$ Pauli bases: the all-$Z$ basis, the $(N-1)$ bases with $XX$ at a pair of
neighbouring sites (with $Z$ elsewhere), and the $(N-1)$ bases with $YY$ at a
pair of neighbouring sites (with $Z$ elsewhere). Note that for the Hamiltonian
given by~\cref{eq:schwinger_h}, the samples provide an estimation of the energy.

The results of the scaling study shown in~\cref{fig:schwinger_errmit_performance}e
and~\cref{fig:schwinger_errmit_performance}f use a modified VQE implementation for
computational efficiency. Instead of evolving the circuit using the entangling
Hamiltonian, we use an entangling layer $\mathcal{O}_N$ comprising two layers of
two-qubit gates simulated without noise, as depicted
in~Extended Data Fig. 1b. The entangling layer was chosen
to exactly preserve the symmetries of the lattice Schwinger model, while being
easier to simulate numerically than evolution with $\hat H_E$. Note that, since
it is composed of nearest-neighbour gates, it is also suited for superconducting
quantum hardware, especially to capacitively coupled, flux-tunable transmon
qubits, where the interaction $XX + YY$ can be easily
implemented~\cite{krantz2019quantum}. For the scaling study, 1024 samples are
taken in each basis to estimate the energy during SPSA optimization. The
hyperparameter $A$ of SPSA is increased to 20 and the other parameters are left
unchanged.

\section*{Acknowledgements}

E.~R.~B., F.~H., B.~K., and P.~R. thank 1QBit for financial support. During part
of this work, E.~R.~B. and F.~H. were students at the Perimeter Institute and
the University of Waterloo and received funding through Mitacs, and F.~H. was
supported through a Vanier Canada Graduate Scholarship. Research at the
Perimeter Institute is supported in part by the Government of Canada through the
Department of Innovation, Science and Economic Development and by the Province
of Ontario through the Ministry of Colleges and Universities. E.~R.~B. was also
supported with a scholarship through the Perimeter Scholars International
program. P.~R. thanks the financial support of Mike and Ophelia Lazaridis, and
Innovation, Science and Economic Development Canada. J.~C. acknowledges support
from the Natural Sciences and Engineering Research Council, a Canada Research
Chair, the Shared Hierarchical Academic Research Computing Network, Compute
Canada, a Google Quantum Research Award, and the Canadian Institute for Advanced
Research (CIFAR) AI chair program. Resources used by J.~C. in preparing this
research were provided, in part, by the Province of Ontario, the Government of
Canada through CIFAR, and companies sponsoring the Vector Institute
(\url{www.vectorinstitute.ai/#partners}). P.~R. thanks Christine~Muschik for
useful conversations. The authors thank Marko Bucyk for carefully reviewing and
editing the manuscript.

\section*{Author Contributions}

E.~R.~B. and F.~H. made equal contributions. They developed the codebase for all
studies, performed numerical experiments, and analyzed the results. E.~R.~B.
performed experiments using the IBM quantum processor. E.~R.~B. and F.~H.
focused on the quantum chemistry and the lattice Schwinger model case studies,
respectively. All authors contributed to ideation and dissemination. J.~C. and
P.~R. contributed to the theoretical foundations and design of the method.

\section*{Data availability} The experimental and numerical quantum simulation
 measurement data are available from 
\cref{fig:quantum_chemistry_error_mitigation_performance} for H$_2$, LiH 
as well as the measurement data used in \cref{fig:schwinger_errmit_performance}a
through \cref{fig:schwinger_errmit_performance}d for the 8 site lattice Schwinger 
model at \url{https://github.com/1QB-Information-Technologies/NEM} 
(see Zenodo repository~\cite{bennewitz2022nemsoftware}).

\section*{Code availability statement} The numerical implementation 
of neural error mitigation as well as the code used to numerically implement the 
quantum simulations studied in the manuscript can be found at 
\url{https://github.com/1QB-Information-Technologies/NEM} 
(see Zenodo repository~\cite{bennewitz2022nemsoftware}).

\section*{Competing Interests Statement}
The authors declare no completing interests.


%
\clearpage
\onecolumngrid

\makeatletter
\def\H2{H$_2$}

\setcounter{section}{0}
\setcounter{equation}{0}
\setcounter{figure}{0}
\setcounter{table}{0}
\setcounter{page}{1}

\makeatletter

\renewcommand{\theequation}{S\arabic{equation}}
\renewcommand{\thefigure}{S\arabic{figure}}
\renewcommand{\thetable}{S\arabic{table}}
\renewcommand{\bibnumfmt}[1]{[#1]}
\renewcommand{\citenumfont}[1]{#1}

\begin{center}

\textbf{\large Supplementary Information:\\
Neural Error Mitigation of Near-Term Quantum Simulations}
\end{center}

\section{Details of the Transformer Neural Quantum State}
\label{sec:app:transformer}

\begin{wrapfigure}{r}{0.3\textwidth}
\includegraphics[width=\linewidth]{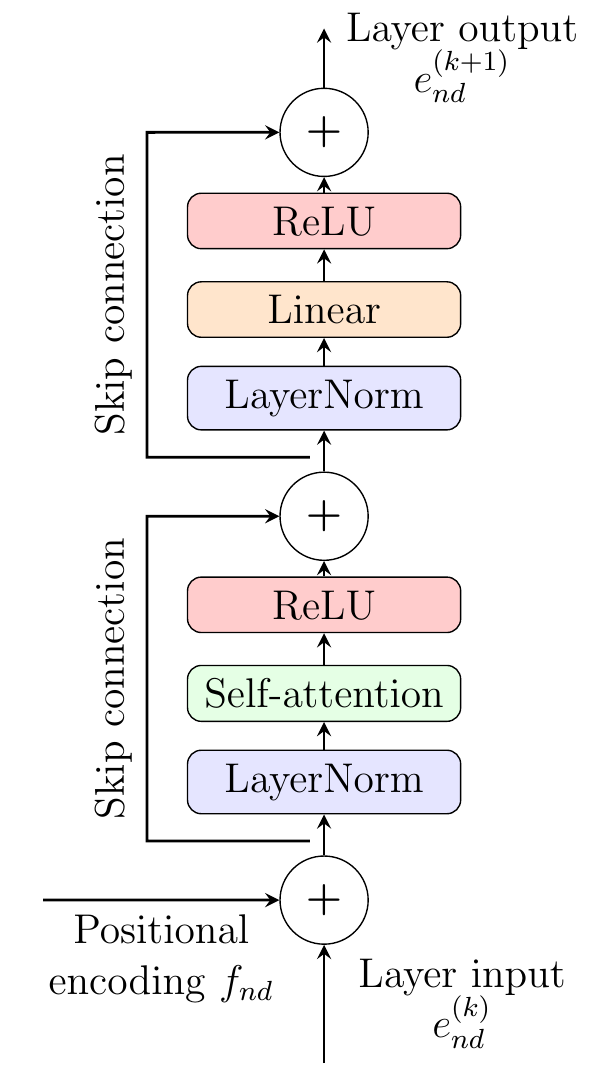}
\caption{\textbf{Illustration of a Transformer layer $|$} The Transformer
consists of several layers acting in sequence. The central components of a
Transformer layer are the self-attention block and linear layer. To enable a
position-aware output, at each layer a positional encoding of the index $n$ is
added to the input. To enhance stability, a skip connection and layer
normalization are employed.}
\label{fig:transformer_schematic}
\end{wrapfigure}

We use the Transformer~\cite{vaswani2017attention} neural network architecture
to represent neural quantum states. The Transformer can be applied to  represent
probability distributions and other functions over one-dimensional sequences of
discrete data. At its core is self-attention, a mechanism that learns
correlations by assigning attention weights to each pair of positions (i.e., how
much the output at one position of a given sequence should depend on the input
at another position). Each layer of the Transformer contains a self-attention
component, which, in turn, consists of several independent so-called
self-attention heads. Within a self-attention component, each of these heads
have independent parameters, and act on the layer inputs independently in
parallel. The  output of the self-attention component is the concatenation of
the outputs of each  head. In each head, for every pair of positions in the
sequence, the attention weight is computed from the inputs at those positions,
using  key and query vectors. The output at  each position is a sum, weighted by
the attention weights, of the value vectors at all positions. To improve the
representation power of the model, the self-attention components alternate with
linear  layers, which act on each position individually. To ensure the stability
of the model, skip-connections and  layer normalization are employed within each
layer. A single Transformer layer with all of its components is depicted
in~\cref{fig:transformer_schematic}.

Recall from the Methods section of the main manuscript that the  neural quantum
state maps a bitstring $s \in \{0,1\}^N$ to a tuple $(p(s), \varphi(s))$, with
$\langle s|\psi\rangle = \sqrt{p(s)} e^{i\varphi(s)}$, where $N$ is the number of qubits.
The computational cost of calculating $\langle s | \psi \rangle$ for a single computational basis state
 $s$ is quadratic in the number of qubits, and alternative versions of the Transformer architecture
  improve this cost to being linear (e.g.,~\cite{wang2020linformer}).

More specifically, the
outputs $p(s)$ and $\varphi(s)$ are obtained from the bitstring $s$ as follows.
The bitstring $s$ is extended to $\tilde s = (0, s)$ by prefixing a zero bit.
Each bit $\tilde s_n$ of the input is encoded into a $D$-dimensional
representation space using a learned embedding, yielding the encoded bit
$e^{(0)}_{nd}$ with $n \in \{0, \ldots, N\}$ and $d \in \{1, \dots, D\}$. Each
index $n$ is encoded using a learned positional encoding, yielding the encoded
index $f_{nd}$. The encoded input is processed using $K$ identical layers. Each
layer consists of a masked multi-head self-attention
mechanism~\cite{vaswani2017attention}, with $H$ heads, followed by a linear
layer, with skip-connections and layer normalization. The weights of the
attention matrices and the linear layer are shared between the positions $n$,
but not between the layers $k$.

The parameters of a single layer $k$ are:
\begin{enumerate}
\item The query, key, and value matrices $Q^{(k)}_{hid}$, $K^{(k)}_{hid}$, and
$V^{(k)}_{hid}$. The index $h \in \{1, \ldots, H\}$ labels the  self-attention
heads. The index $i \in \{1, \ldots, D/H\}$ runs over a representation space of
dimension $D/H$, where we require that $D/H$ be an integer. As before, we have
$d \in \{1, \ldots, D\}$.

\item A matrix to process the output of the self-attention heads,
$O^{(k)}_{de}$, with $d, e \in \{1, \ldots, D\}$.

\item A weight matrix and a bias vector of the linear layer, $W^{(k)}_{de}$ and
$b^{(k)}_d$, with $d, e \in \{1, \ldots, D\}$.
\end{enumerate}
The self-attention component acts as follows:
\begin{enumerate}

\item We denote the input to the component by $i^{(k)}_{\text{SA},nd}$, where $n
\in \{0, \ldots, N\}$ and $d \in \{1, \ldots, D\}$. Query, key, and value
vectors are computed as $q^{(k)}_{nhi} = \sum_d Q^{(k)}_{hid}
i^{(k)}_{\text{SA},nd}$, $k^{(k)}_{nhi} = \sum_d K^{(k)}_{hid}
i^{(k)}_{\text{SA},nd}$, and $v^{(k)}_{nhi} = \sum_d V^{(k)}_{hid}
i^{(k)}_{\text{SA},nd}$.

\item Attention weights are computed as follows. Compute the attention scores
$s^{(k)}_{nmh} = \sum_i q^{(k)}_{nhi} k^{(k)}_{mhi}$. Mask the attention scores
by setting $s^{(k)}_{nmh} = -\infty$ whenever $m < n$. Compute the attention
weights using the softmax function given by
\begin{equation}
  w^{(k)}_{nmh} = \frac{\exp(s^{(k)}_{nmh})}{\sum_{m'} \exp(s^{(k)}_{nm'h})}.
\end{equation}
The masking ensures that $w^{(k)}_{nmh} = 0$ whenever $m < n$.

\item The output of each attention head is $a^{(k)}_{nhi} = \sum_m w^{(k)}_{nmh}
v^{(k)}_{mhi}$. At each position $n$, concatenate the output of the attention
heads; that is, reshape the indices $h$ and $i$ into one index $d$, giving
$a^{(k)}_{nd}$.

\item The output of the self-attention component is $o^{(k)}_{\text{SA},nd} =
\sum_e O^{(k)}_{de} a^{(k)}_{ne}$. Due to the masking, the output at position
$n$ depends on the inputs only at positions $m \le n$. We write the action of
the entire self-attention component more abstractly as $A^{(k)}$, that is,
$o^{(k)}_\text{SA} =A^{(k)}(i^{(k)}_\text{SA})$.

\end{enumerate}

The linear component acts on an input $i^{(k)}_{\text{L},nd}$ as
$o^{(k)}_{\text{L},nd} = \sum_e W^{(k)}_{de} i^{(k)}_{\text{L},nd} + b^{(k)}_d$,
which we write more abstractly as  $o^{(k)}_{\text{L}} =
L^{(k)}(i^{(k)}_{\text{L}})$. Both components are wrapped with a
skip-connection, layer normalization \cite{ba2016layer}, and a ReLU activation
function. The output of the wrapped self-attention component $A^{(k)}$, acting
on an input $i^{(k)}_{nd}$, is
\begin{equation}
  a^{(k)} = G(A^{(k)}, i^{(k)})
  = i^{(k)} + \text{ReLU}\Big(A^{(k)}\big(\text{LayerNorm}(i^{(k)})\big)\Big),
\end{equation}
where $\text{ReLU}(x) = \max(x, 0)$ is the ReLU activation function acting
component-wise, and layer normalization is applied on the last dimension of
$i^{(k)}_{nd}$. We have introduced the notation $G$ for the wrapping. The linear
component $L^{(k)}$ is wrapped in the same manner, giving $e^{(k)} = G(L^{(k)},
a^{(k)})$. In sum, the action of the entire Transformer layer is as follows:
\begin{enumerate}
\item The input to the $k$-th layer is $i^{(k)}_{nd} = e^{(k-1)}_{nd} + f_{nd}$.

\item The wrapped self-attention component is applied to give
$a^{(k)} = G(A^{(k)}, i^{(k)})$.

\item The wrapped linear layer is applied to give
$e^{(k)} = G(L^{(k)}, a^{(k)})$.
\end{enumerate}
After the final Transformer layer, the outputs of the neural quantum state are
obtained from the final representation $e^{(K)}_{nh}$ as follows. Scalar-valued
logits $\ell_n$ are obtained from $e^{(K)}_{nh}$ using a linear layer, with
weights shared among different positions $n$. The logits are used to obtain
conditional probabilities according to

\begin{align}
  p(s_n = 1| s_1, \ldots, s_{n-1})
    ={}& \sigma(\ell_{n-1})\qquad\text{and}\nonumber\\
  p(s_n = 0| s_1, \ldots, s_{n-1})
    ={}& 1-p(s_n= 1| s_1, \ldots, s_{n-1})
    =\sigma(-\ell_{n-1}), \qquad\text{where}~n \in \{1, \ldots, N\},
\end{align}
and $\sigma(\ell_{n-1})= \tfrac{1}{1+e^{-\ell_{n-1}}}$ is the logistic sigmoid
function. The conditional probabilities give $p(s)$ via the conditional
probability chain rule

\begin{align}
  p(s) = \prod_{n=1}^N p(s_n | s_1, \ldots, s_{n-1}),
\end{align}
where $p(s)$ is an automatically normalized probability distribution.

The factorization of the probabilities $p(s)$, along with the fact that the
neural network output at position $n$ depends only on the positions $m \le n$,
may be leveraged to efficiently draw unbiased samples from the probability
distribution $p$. To do so, we proceed a single bit at a time. First, we compute
$p(s_1 | s_0 = 0)$, and sample the bit $s_1$ from the resulting probability
distribution. We then compute $p(s_2 | s_0, s_1)$ and sample the bit $s_2$ from
it, and so on, until all bits have been sampled. The sampling is needed when
training the NQS using VMC, as explained in the next section.

The phase $\varphi(s)$ is obtained by forming a vector $E^{(K)}=(e^{(K)}_0,
\ldots, e^{(K)}_N)$ by concatenating the final representations, and projecting
$E^{(K)}$ to a scalar value using a linear layer. Our Transformer is implemented
in PyTorch~\cite{paszke2019pytorch}, and is partially inspired by aspects of the
implementations in Refs.~\cite{dai2019transformer, parisotto2020stabilizing}.

\section{VMC Training and Regularization}
\label{app:vmc}

In the final step of NEM described in the Methods section in the main
manuscript, the neural quantum state is trained using VMC. We add the regularizer
\begin{equation}
  L_\text{reg} = -\epsilon_\text{reg}
  \sum_{s\in\{0,1\}^N} \Big| \langle s | \psi_{\vec\lambda}\rangle \Big|
\end{equation}
to the loss function, where $\epsilon_\text{reg}$ is a coefficient that is
decreased over the course of training. The regularizer maximizes the $L_1$-norm of the
wavefunction, thereby discouraging sparsity in the early stages of  training. 
For ground states that are known to overlap greatly with one or a
 few ground states, this regularization technique encourages the NQS to capture 
 subdominant amplitudes rather than collapsing onto the dominant contributions
  (e.g., the Hartree--Fock states, or the lattice Schwinger ground states at $m \rightarrow \pm \infty$) early in the training process.

As with the energy, the regularizer and
its gradient with respect to the parameters $\vec\lambda$ of the neural quantum
state $|\psi_{\vec\lambda}\rangle$ need to be estimated from samples using the
Monte Carlo method, giving the expressions

\begin{align}
  L_\text{reg}
  ={}& -\epsilon_\text{reg} \mathbb{E}_{s \sim p_{\vec\lambda}}
  \Big[\Big| \langle s | \psi_{\vec\lambda} \rangle \Big|^{-1} \Big]
  \approx -\epsilon_\text{reg} \frac{1}{b}
  \sum_{i=1}^b\Big| \langle s^i | \psi_{\vec\lambda}
  \rangle \Big|^{-1}\qquad \text{and} \nonumber\\
  \nabla_\theta L_\text{reg}
  ={}& - \epsilon_\text{reg} \mathbb{E}_{s \sim p_{\vec\lambda}}
  \Big[ \Big| \langle s | \psi_{\vec\lambda}
  \rangle \Big|^{-1} \nabla_{\vec\lambda}
  \Re(\ln \langle s | \psi_{\vec\lambda} \rangle)\Big]
  \approx - \epsilon_\text{reg} \frac{1}{b}
  \sum_{i=1}^b \Big[\Big| \langle s^i | \psi_{\vec\lambda}
  \rangle \Big|^{-1} \nabla_{\vec\lambda}
  \Re(\ln \langle s^i | \psi_{\vec\lambda} \rangle)\Big].
\end{align}
Here, $\Re$ denotes the real part, the expectation values are taken over
$p_{\vec\lambda}(s) = |\langle s| \psi_{\vec\lambda}\rangle|^2$, $s^i$ are
samples from the same distribution, and $b$ is the batch size used in VMC.

The regularizer's gradient is added to the energy's gradient at every iteration.
For completeness, we reproduce the VMC algorithm here:
\begin{enumerate}

\item Draw $b$ samples $\{s^1, \ldots, s^b\}$ from $p_{\vec\lambda}(s) =
\|\langle s | \psi_{\vec\lambda} \rangle |^2$.

\item Compute the local Hamiltonians for all $s^i \in \{s^1, \ldots, s^b\}$:
\begin{equation}
  H_\text{loc} (s^i)
  = \mathlarger
  \sum_{\substack{t \in \{0,1\}^N \\ \langle s^i|\hat H |t\rangle \neq 0}}
  \frac{\langle s^i | \hat H | t \rangle \langle t | \psi_{\vec\lambda} \rangle}
  {\langle s^i | \psi_{\vec\lambda} \rangle}
\end{equation}

\item Estimate the energy and its gradient with respect to the parameters
${\vec\lambda}$ of $\ket{\psi_{\vec\lambda}}$ using

\begin{align}
  E \approx{}&
  \frac{1}{b} \Re\Big(\sum_{i=1}^b H_\text{loc} (s^i) \Big)
  \qquad \text{and} \nonumber\\
  \nabla_\theta E \approx{}&
  \frac{2}{b} \Re\Big[ \sum_{i=1}^b (E_\text{loc} (s^i) - E)^*
  \nabla_\theta \ln \langle s_i | \psi_{\vec\lambda} \rangle\Big],
\end{align}
where $\ln \langle s |\psi_{\vec\lambda} \rangle = \ln \sqrt{p(s)} + i
\varphi(s)$ in terms of the neural network's outputs.

\item Estimate the regularizer and its gradient.

\item Add the energy's gradient and the regularizer's gradient, and update the
parameters using the Adam optimizer.

\end{enumerate}

\section{Experimental VQE Results for L\lowercase{i}H}

\begin{figure}
\centering
\includegraphics[width=0.65\linewidth]
{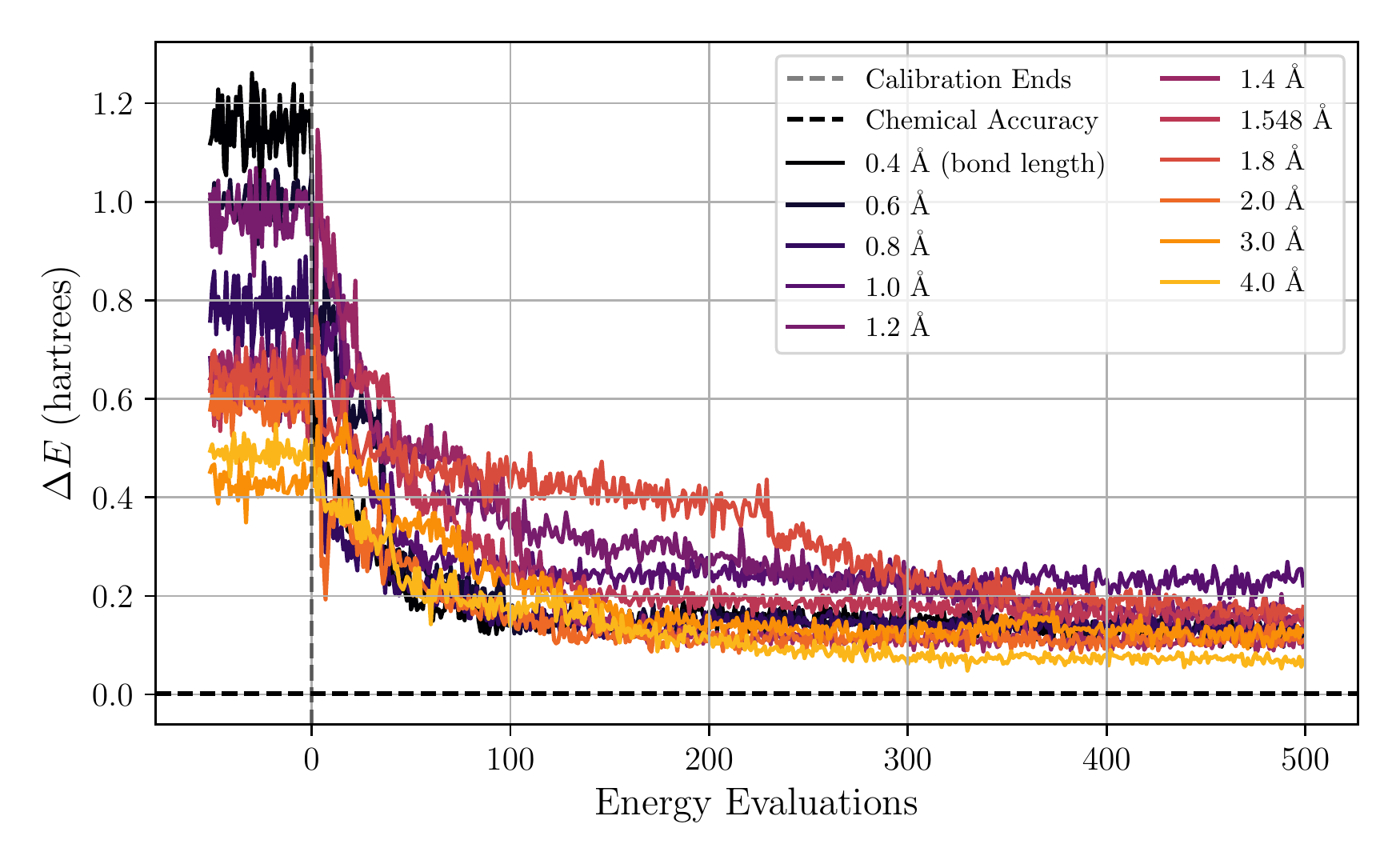}
\\[-3ex]
\caption{\textbf{Experimental VQE energy optimization curves $|$} Energy
evaluations during SPSA optimization for LiH ground states at different bond
lengths using a five-qubit superconducting quantum computer. In order to compare
optimization curves for all bond lengths, we plot the energy difference to the
ground state as a function of SPSA energy evaluations. Before optimization, the
energy is estimated 25 times and used to calibrate the initial step size of the
algorithm. These energy values are reported to the left of the vertical dashed
line.  During each SPSA iteration, the energy is estimated using 25 measurement
bases with 1024 measurements per measurement basis. The parameters $\vec \theta$
are updated 250 times, resulting in 500 energy evaluations.  Note that three
bond lengths show incomplete convergence by the end of the optimization process,
with one bond length (1.8), still decreasing. Although VQE does not reach
convergence for all bond lengths, neural error mitigation is able to improve the
results up to chemical accuracy, with low infidelity, as shown in the top row
of~Fig. 2 in the main manuscript.}
\label{fig:quantum_chemistry_ibmq_rome_optimization}
\end{figure}

We can analyze the quality of the variational procedure implemented on the
five-qubit IBMQ-Rome device by looking at the energy optimization curves of the
experimental  results of VQE implemented for various bond lengths of LiH
(see~\cref{fig:quantum_chemistry_ibmq_rome_optimization}). Generally, the
optimization curves show that the experimental VQE procedure performs as
expected, with energy optimization curves decreasing and converging to low
energy estimates. However, we also see that the simulation is unable to reach
high-accuracy energy estimates, $\Delta E \lessapprox 10^{-3}$. This is due to
the effects of noise on, and limitations of, the variational quantum
circuit~\cite{kandala2017hardware}. Despite these errors, neural error
mitigation is able to extend all experimental results up to chemical accuracy,
with low infidelities. We refer the reader to~\cref{tab:hyperparameters} for the
specific hyperparameters used in implementing VQE and NEM.

\section{Quantum State Reconstruction}

\begin{figure}
  \begin{subfigure}{0.45\linewidth}
    \centering
    \caption{}
    \includegraphics[width=\linewidth]
    {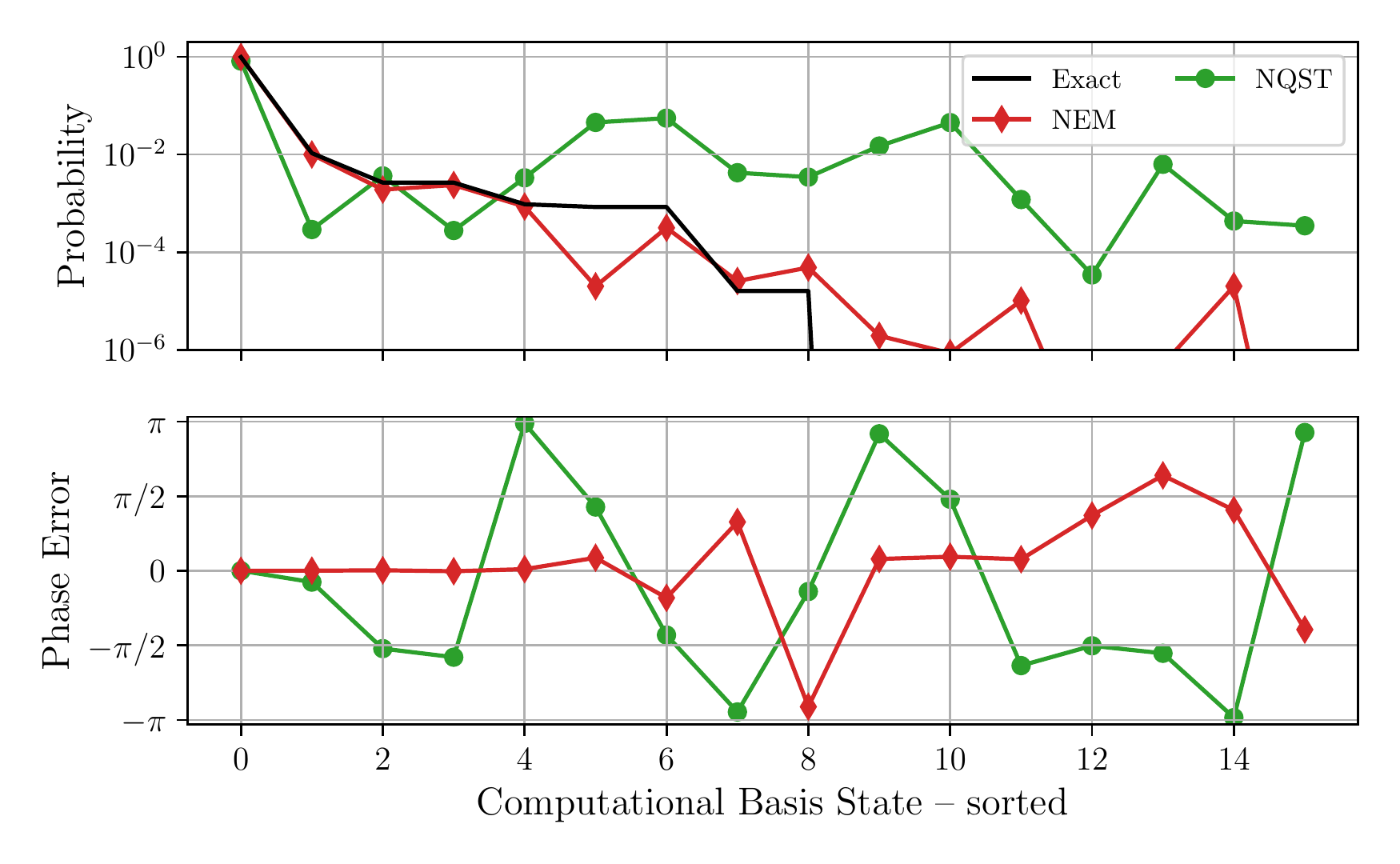}
    \label{}
    \label{fig:lih_quantum_states_experimental}
  \end{subfigure}
  \hspace{10pt}
  \begin{subfigure}{0.45\linewidth}
    \centering
    \caption{}
    \includegraphics[width=\linewidth]
    {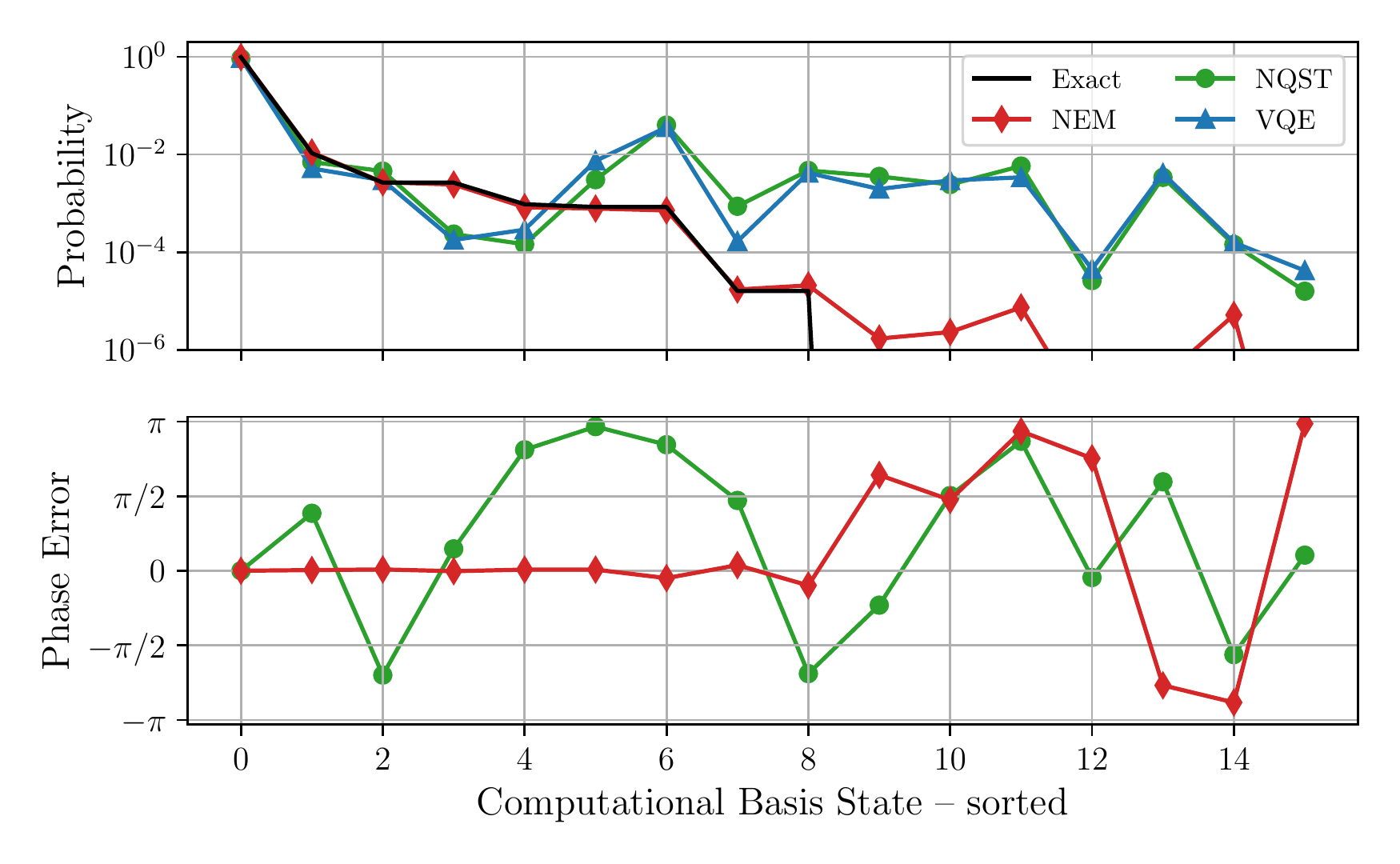}
    \label{fig:lih_quantum_states_simulated}
  \end{subfigure}
  \\[-3ex]
\caption{\textbf{Quantum states at each step in the neural error mitigation
algorithm for LiH $|$} The quantum states obtained by VQE (blue triangles),
neural quantum state tomography (green circles), and neural error mitigation
(red diamonds) for the LiH ground state at a bond length of $l=1.4$~\r{A}. In
contrast to the neural quantum state, the VQE state is given by a density matrix
representing a mixed state because VQE is simulated using a noisy density matrix
simulator. Whereas the probability distribution is given by $p(s) =$ Tr$(\rho
\ket{s}\bra{s})$, the phase for a mixed state is not well-defined, and therefore
not reported. In each subfigure, the top panel shows the probabilities for the
NQS given by $p(s) = |\langle s|\psi \rangle|^2$ for each computational basis
state $|s \rangle$. Computational basis states are sorted according to these
probability amplitudes. The bottom panel shows the phase error of the NQS
relative to the ground state given by $\arg (\langle s | \psi \rangle) - \arg
(\langle s | \psi_g \rangle)$, where the global phase is fixed to a phase error
of zero for the computational basis state that has the largest ground state
probability. Neural error mitigation applied to an experimentally prepared VQE
result is shown on the left~\subref{fig:lih_quantum_states_experimental}, and
NEM applied to a numerically simulated VQE result is shown on the
right~\subref{fig:lih_quantum_states_simulated}. Under our qubit encoding for
the electronic structure problem, the Hartree--Fock state is mapped to a
computational basis state and constitutes the dominant contribution to the exact
ground state corresponding to the 0-th state on the horizontal axis. This
results in an exact probability distribution that has a sharp peak, as shown
traced by the black line. We see that the neural quantum states trained using
NQST approximately reconstruct the VQE quantum state's probability distribution,
for the numerically simulated results. In addition, the final probability
distribution represented by the NEM states very accurately represents the basis
states with exact probabilities greater than $10^{-5}$ for numerical simulations
and greater than $10^{-4}$ for our experimental results. The phase errors are
shown in the bottom panel, where we see that the final NEM quantum state
accurately reconstructs the phase for computational basis states with
probabilities greater than $10^{-5}$ for numerical simulations and greater than
$10^{-4}$ for our experimental results.}
\label{fig:lih_quantum_states}
\end{figure}

\begin{figure}
\includegraphics[width=0.6\linewidth]{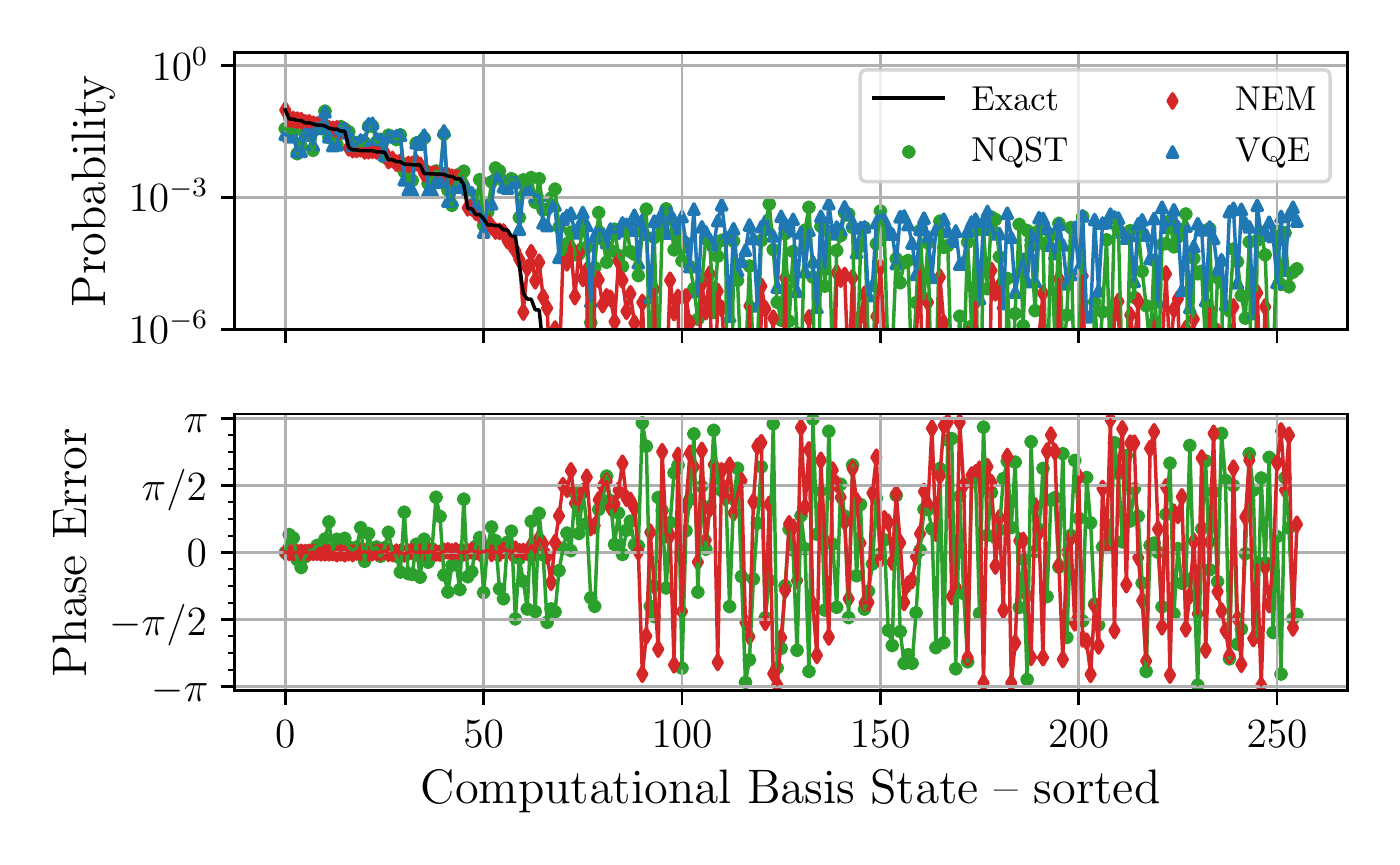}\\[-3ex]
\caption{\textbf{Quantum states at each step of the neural error mitigation
algorithm for the \textbf{$N=8$} lattice Schwinger model $|$} The quantum states
obtained by VQE (blue triangles), by NQST (green circles), and after completion
of NEM (red diamonds) are shown, for the lattice Schwinger model at $m=-0.7$.
Similar to \cref{fig:lih_quantum_states}, the top panel shows the probabilities
of each basis state sorted according to their probability amplitudes, and the
bottom panel shows the phase errors with respect to the exact ground-state
phase. In contrast to the NQS results, the VQE state is given by a mixed-state
density matrix. Therefore, we report the VQE probability distribution but not
the phase. Due to the symmetries present in the lattice Schwinger ground state,
many computational basis state amplitudes are zero in the exact ground-state
probability distribution (solid black line). While the probabilities of the VQE
state somewhat follow the ground state, the VQE state overestimates many of the
computational basis states' probabilities that do not contribute to the exact
ground state. This is explained by the effects of noise and because the quantum
circuit employed only approximately preserves the symmetries of the model. We
can see that NQST successfully reconstructs the VQE state, accurately modelling
the probability distribution for the computational basis states that contribute
to the exact ground state. Additionally, NQST provides an estimate of the phase,
showing a phase error  confined mostly to the range $[-\pi/2, \pi/2]$ for the
computational basis states that make non-zero contributions to the exact ground
state. At the end of NEM, the final probability distribution represented by the
NEM state accurately represents the probabilities and phases of the
computational basis states that have exact probabilities greater than
$10^{-5}$.}
\label{fig:schwinger_state}
\end{figure}

An important feature of NEM is the ability to analyze the quality of the
reconstructed neural quantum state at various stages of the NEM algorithm. Since
NEM employs NQS at the core of its construction, we have direct access to the
reconstructed quantum state through the probability, $p(s)$, and phase,
$\phi(s)$, distributions given by, $p(s) = |\braket{s | \psi}|^2$ and $\psi(s) =
\arg(\braket{s|\psi})$. We analyze the reconstructed quantum state obtained
after NQST as well as the final NEM procedure for both of the systems studied in
this paper, and compare the probability distributions modelled by the neural
quantum states to the VQE result. The VQE state is given by a density matrix
$\rho$, because VQE is numerically simulated using a density matrix simulator.
The probability amplitudes of $\rho$ are given by $p(s) =$ Tr$(\rho
\ket{s}\bra{s})$. Since we simulate VQE with noise, the VQE density matrix
$\rho$ describes a mixed state instead of a pure state. For mixed states, the
phase is not well-defined, and therefore not reported
in~\cref{fig:lih_quantum_states} and~\cref{fig:schwinger_state}.

In~\cref{fig:lih_quantum_states}, we show the quantum states estimated at each
stage of our process for the LiH ground states at a bond length of
$l=1.4$~\r{A}, prepared both
experimentally~(\cref{fig:lih_quantum_states_experimental}) and numerically
~(\cref{fig:lih_quantum_states_simulated}). We see that, for numerical results,
where we have access to the VQE quantum state, neural quantum state tomography
accurately reconstructs the optimized VQE state using the chosen measurement
bases.  For the experimental results, where we do not have access to the final
VQE quantum state, the neural quantum state trained using neural quantum state
tomography acts as an estimator of the final state expressed by the quantum
device. At this stage, the neural quantum state trained using NQST has not
captured the exact ground state's phase structure  and inaccurately represents
some of the non-dominant computational basis states. After the NEM algorithm has
been completed, the final NEM state achieves accurate representations of both
the probability distribution and phase for the computational basis states whose
exact probabilities are greater than $10^{-5}$ (or greater than $10^{-4}$ for
experimental results). In the process of improving the energy estimation
achieved by VQE, NEM reconstructs and improves the ground-state wavefunction
itself. From another perspective, the classical ansatz trained through this
process extends the ``lifetime" of the quantum
simulation~\cite{torlai2018neural}, allowing for its use in future work, without
having to repeat the experiment.

Figure~\ref{fig:schwinger_state} shows a representation of the VQE state, as
well as the neural quantum state after applying NQST and after having completed
NEM for the lattice Schwinger model. The phase structures of the lattice
Schwinger ground states follow a sign rule and have real amplitudes. Although
possible, we do not enforce the sign rule or any symmetries in our neural
quantum states in order to show the general applicability of NEM.  While the
fidelity of the NQST state with respect to the ground state is only  0.71, the
errors in the complex phases that correspond to the computational basis states
with non-zero amplitudes are relatively small, and mostly confined to the  range
$[-\frac\pi2, \frac\pi2]$. Given that converging to an accurate phase structure
is one of the main difficulties encountered in training a neural network using
VMC~\cite{szabo2020neural}, the NQST state may provide a good starting  point
from which it could be easy for the VMC algorithm to converge to a good
approximation of the ground state. After the NEM algorithm has completed, the
final NEM state achieves an accurate representation of both the probability
distribution and the phases of the lattice Schwinger ground state, specifically
for the computational basis states whose exact probabilities are greater than
$10^{-5}$.

\section{Comparison of NEM to VMC}
\label{sec:app:vmc-comparison}

The key observation outlined in this paper is that NQST and VMC can be combined
to form a post-processing error-mitigation strategy for ground-state preparation
when the two procedures are conducted using a common neural network ansatz. In
addition to analyzing how well NEM improves the results obtained from noisy
quantum simulations, it is also useful to compare NEM to its classical
counterpart: training a neural quantum state using only the variational Monte
Carlo algorithm, hereafter referred to as \textit{standalone} VMC. We compare
the performance of both methods as a function of VMC batch size, which is the
number of samples used in estimating the gradient and updating the neural
network's parameters during VMC~(see~\cref{app:vmc}). Given that we fix the
total number of iterations used in training, the batch size is indicative of the
classical computational resources required. Note that increasing the batch size
decreases the variance of the energy's gradient estimate. For larger systems,  a
large batch size is often required in order to reach chemical accuracy, which
imposes a bottleneck on the possible applications of
VMC~\cite{choo2020fermionic}. Another way to increase the amount of classical
computational resources, which may exhibit different scaling, would be to
increase the number of iterations at a fixed batch size.

\begin{figure}[t]
\begin{subfigure}{0.45\linewidth}
  \subcaption{}
  \includegraphics[width=.99\linewidth]{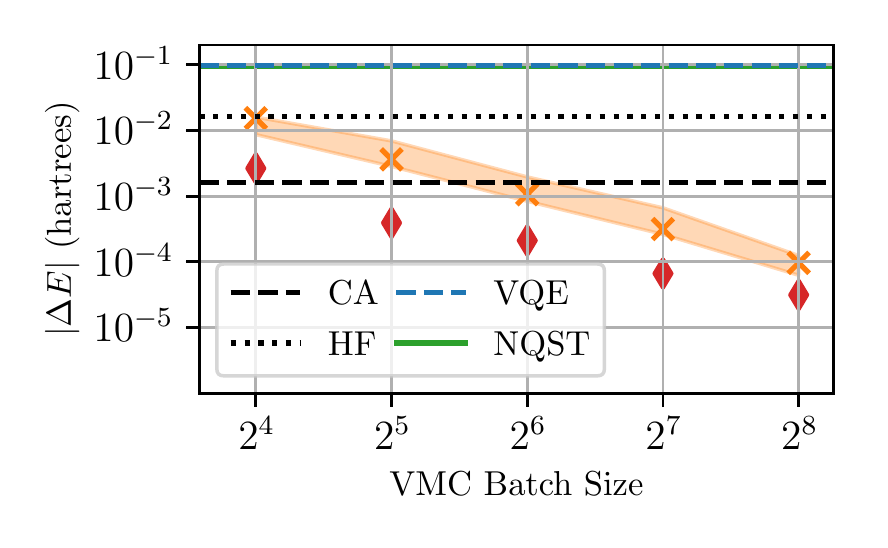}
  \label{subfig:lih_batch_size_analysis_energy}
\end{subfigure}
\begin{subfigure}{0.45\linewidth}
  \subcaption{}
  \includegraphics[width=.99\linewidth]{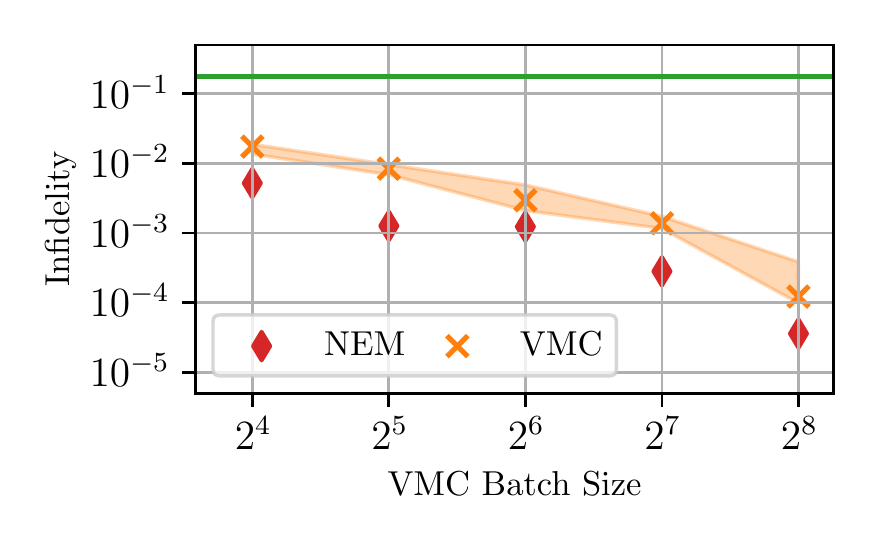}
  \label{subfig:lih_batch_size_analysis_infidelity}
\end{subfigure}\\[-5ex]
\caption{\textbf{Comparison between neural error mitigation and variational
Monte Carlo for quantum chemistry $|$} The performance of neural error
mitigation (red diamonds) on experimental VQE results compared to the
performance of training a neural quantum state  using standalone VMC (orange
crosses) is shown. Both methods are compared as a function of the VMC batch size
for the LiH ground state for $l = 1.4$. Both methods are performed using a
neural quantum state that has both the same architecture hyperparameter values.
The median results for VMC are shown for 10 runs, and the shaded regions show the interquartile
range.
In~\subref{subfig:lih_batch_size_analysis_energy}, the chemical accuracy (CA)
and Hartree--Fock (HF) energy error are reported.   For LiH, NEM applied to the
experimental results outperforms standalone VMC, achieving chemical accuracy at
a batch size at a factor of two earlier than standalone VMC.}
\label{fig:lih_batch_size_analysis}
\end{figure}

In~\cref{fig:lih_batch_size_analysis} we compare the energy error and infidelity
of NEM performed on the experimental VQE result and standalone VMC, for the LiH
molecule at a bond length of $l=1.4$ \r{A} (the same experimental VQE data is
presented in the top row
of Fig. 2 in the main
manuscript). For NEM, we fix the outcome of the first stage of NEM (i.e., the
neural quantum state trained via NQST) and then train the VMC component of NEM
using different batch sizes to investigate how the energy error and infidelity
of the final NEM state scales. The results show that NEM performed using the
experimental results provides an advantage over using only VMC. NEM achieves
chemical accuracy using a lower batch size than VMC. In addition we note that
NEM requires a smaller regularizer compared to conducting the training using
only VMC. Note that the ground states of LiH, a molecular system that can be
mapped to four qubits, can be feasibly solved using current classical methods.

\begin{figure*}
\includegraphics[width=.99\linewidth]{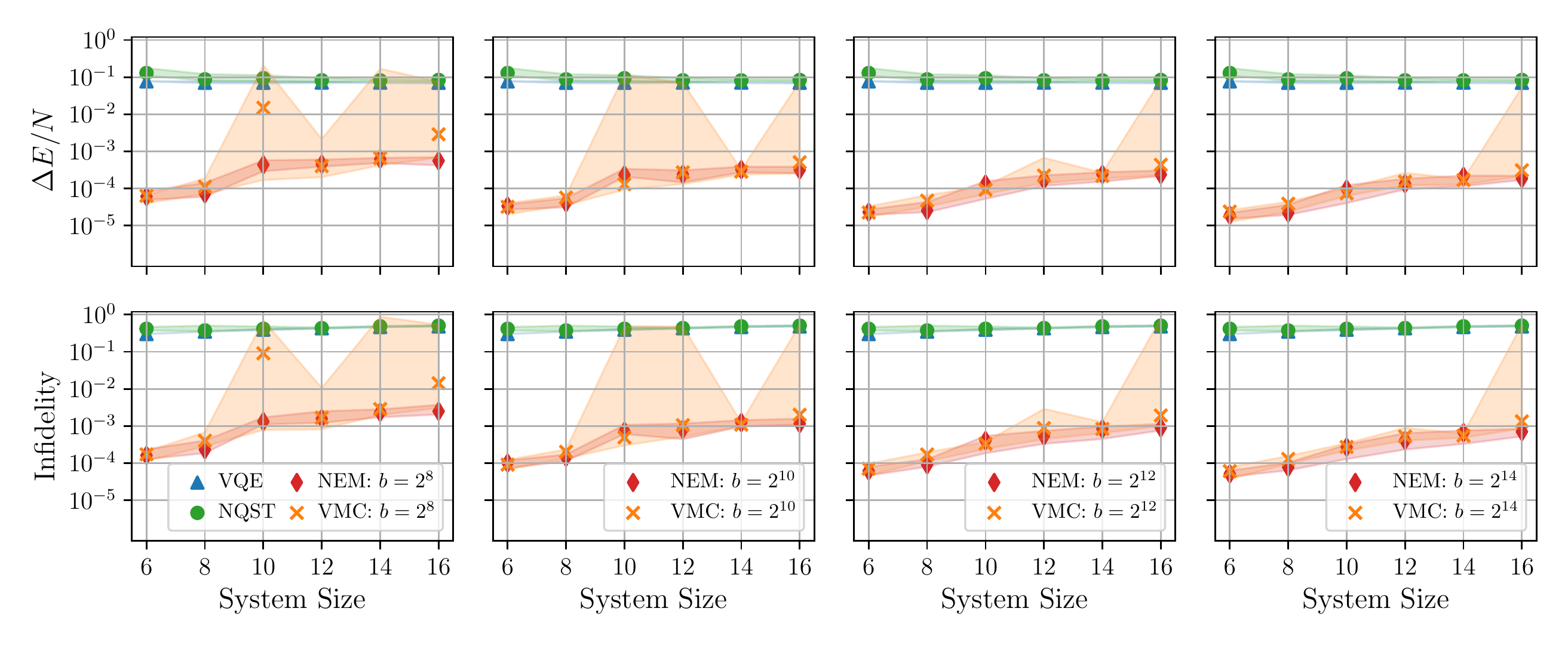}\\[-3ex]
\caption{\textbf{Comparison between neural error mitigation and variational
Monte Carlo for the lattice Schwinger model $|$}  Performance of standalone VMC
on the lattice Schwinger model, compared to the performance of NEM, for various
system sizes   $N$ and batch sizes $b$. While the best results achieved by
standalone VMC are no worse than the best results achieved by NEM, standalone
VMC does not consistently converge to a high-quality approximation of the ground
state. Neural error mitigation is more stable, especially at smaller batch sizes
and larger system sizes. Each panel shows the median results, and the shaded regions show
the interquartile ranges. The
hyperparameter values used for standalone VMC are the same as  those used in the
scaling analysis shown in~Fig. 3e
and~Fig. 3f of the main manuscript, and are listed
in the column ``$>8$ sites'' of \cref{tab:hyperparameters}.}
\label{fig:schwinger_errmit_vs_vmc}
\end{figure*}

For the lattice Schwinger model, we also study the performance of NEM and
standalone VMC as a function of system size and batch
size~(\cref{fig:schwinger_errmit_vs_vmc}). We show that while the best results
for standalone VMC are comparable to the best results of NEM, standalone VMC has
a lower success rate, especially for larger systems and smaller batch sizes.
Conversely, NEM reliably converges to a good approximation of the ground state.
We speculate that the state found by VQE, which is approximated by NQST,
provides a good initialization for training using VMC, and explains the improved
convergence rate of the NEM algorithm. While we expect that, for system sizes
presented, hyperparameter tuning could potentially improve the results of
standalone VMC, we speculate that the increased stability of NEM over standalone
VMC will persist at larger system sizes.

In order to understand the advantages of using quantum resources in conjunction
with classical methods, future research must be conducted to explore whether
using the NEM algorithm for preparing ground states, such as those for larger
systems in quantum chemistry and lattice theories, converges to classical
representations of quantum states that are outside the reach of standalone VMC.
This speculation stems from the fact that both NEM and standalone VMC train a
neural quantum state using the VMC algorithm, with the difference being that NEM
initializes the VMC algorithm using a classical representation of an
experimentally prepared quantum state. One approach could be to determine
whether the NEM algorithm captures features of the experimentally prepared
quantum state, such as superposition and entanglement, in its initial neural
quantum state representation and whether it retains these features throughout
the classical training process. We also speculate that exploring the loss
landscape of VMC can help to delineate the boundary between classically solvable
ground-state preparation problems and those that require quantum resources. In
other words, we ask whether initializing VMC using a classical representation of
an experimentally prepared quantum state relaxes classical resource requirements
of the VMC algorithm, such as the exponentially large amount of memory needed to
describe high-fidelity ground-state representations. The work presented in this
paper provides a framework for investigating the fundamental differences and
potential synergy between quantum and classical information processing.

\section{ Comparison of NEM across different levels of noise in VQE }

The results presented in this paper have shown that NEM can improve estimates of ground states
 and ground state properties obtained from VQE on noisy devices and in noisy simulations by supplementing
 these noisy VQE results with classical simulation methods. However, for high levels of noise,
 we expect the results obtained from VQE to contain less information about the true ground state. Beyond a certain level of noise, we expect
 that supplementing VQE with NEM should not outperform an NQS trained using purely classical methods such as VMC. 
 In order to determine the regime where the combination of VQE and NEM holds the promise of 
 improvement over purely classical methods, we study the performance of NEM as
 a function of VQE noise levels.

We consider the eight-site Schwinger model, 
at the critical point $m=-0.7$, and compare the performance of NEM for VQE noise rates ranging
from zero noise to complete depolarization. The VQE circuit used is the same as the results shown in Fig. 4a
of the main manuscript, and has an initial state of $|01 \cdots 01\rangle$. After each layer of 
the global entangling operation or single-qubit rotations, a depolarizing channel with a variable
rate $\lambda \in [0,1]$ is applied to each qubit. The hyperparameters for NQST and VMC
are the same as in the main manuscript.

\begin{figure*}
	\includegraphics[width=.99\linewidth]{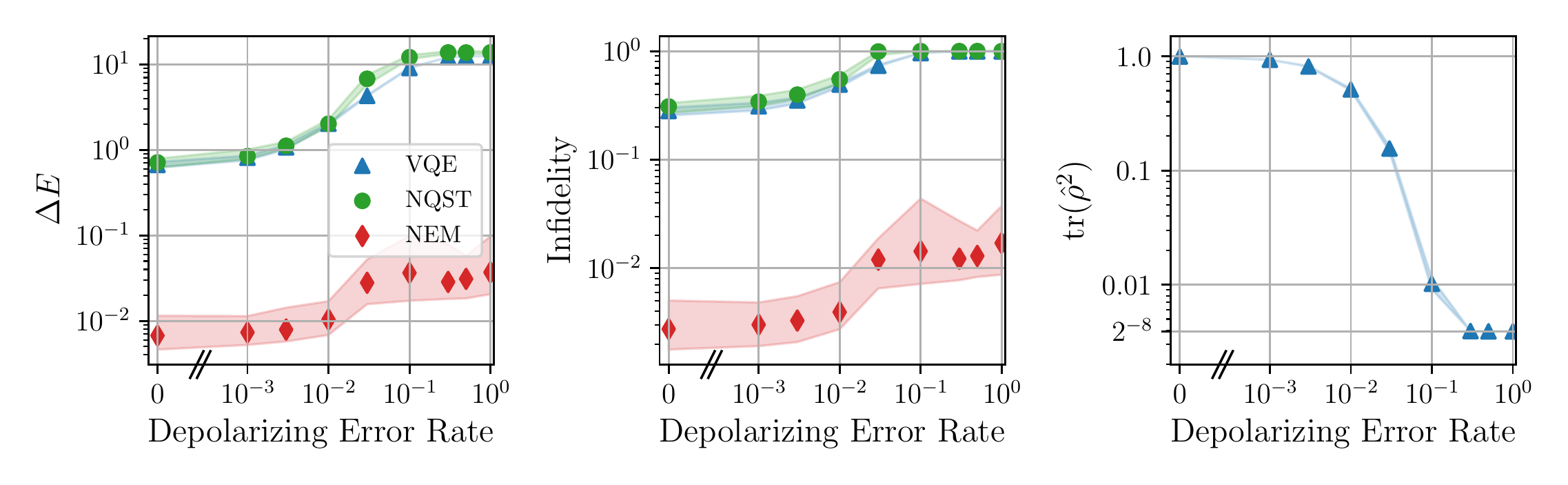}\\[-3ex]
	\caption{\textbf{Comparison of the performance of NEM across different levels of noise in VQE} $|$
	Performance of neural error mitigation applied to VQE results for the eight-site 
	Schwinger model at the critical mass
	$m=-0.7$. Single-qubit depolarizing noise gates of various noise rates $\lambda$ are applied after every
	operation in the simulation of the VQE circuit. 
	The NEM performance when $\lambda = 1$ can be attributed to classical resources only, 
	 since the output of VQE corresponds to a state close to the maximally mixed state.
	At $\lambda \le 10^{-2}$, corresponding to a purity of $\text{tr}(\hat \rho^2) \ge 0.82$, the results 
	of NEM show a clear improvement over those of classical resources alone. Shown is the median
	over 100 runs. The shaded region represents the interquartile range.}
	\label{fig:schwinger_errmit_vs_noise}
\end{figure*}

At a depolarizing error rate of $\lambda=1$, the VQE results are 
highly mixed and thus contain no information about the true ground state or its properties,
 as noise completely dominates the simulation. 
While even in the completely depolarizing simulation, NEM yields an improvement over VQE, its
performance can be attributed to that of VMC which uses only classical computational resources.
Thus, the results obtained when $\lambda=1$ can be used as a benchmark for the performance of standalone VMC. 
The results in \cref{fig:schwinger_errmit_vs_noise} show that at noise rates of
$\lambda = 10^{-2}$ or lower, corresponding to a median purity $\text{tr}(\hat \rho^2) \ge 0.82$ 
of the VQE density matrix, the energy error and infidelity of the NEM
state yield a clear improvement over the NEM results when $\lambda = 1$. 
These results underscore the fact that NEM used in conjunction with VQE shows an improvement over a computation that uses only classical resources given that
the quality of the quantum resources meets a minimum threshold for the system studied. 

\section{Hyperparameter Values}
\label{app:hps}

We present the hyperparameter values  of our numerical studies in Table
\ref{tab:hyperparameters}.

\begin{table}[t!]
\renewcommand{\arraystretch}{1.2}
{\setlength{\extrarowheight}{0.5pt}%
\scriptsize

\begin{tabularx}{\linewidth}
{|p{0.225\linewidth}|p{0.18\linewidth}|p{0.18\linewidth}|X|X|}
  \hline
  & \multicolumn{1}{c|}{\textbf{Lattice}}
  & \multicolumn{1}{c|}{\textbf{Lattice}}
  & \multicolumn{1}{c|}{\textbf{H}$_2$}
  & \multicolumn{1}{c|}{\textbf{LiH}} \\
  & \multicolumn{1}{c|}{\textbf{Schwinger}}
  & \multicolumn{1}{c|}{\textbf{Schwinger}} & & \\
  & \multicolumn{1}{c|}{(8 sites)}
  & \multicolumn{1}{c|}{($> 8$ sites)} & & \\ [2ex]

  \hline
  \multicolumn{5}{c}{\textbf{Variational Quantum Eigensolver}} \\ [1ex]
  \hline

  Iterations & 200 & 200 & 250 & 250 \\

  Post-rotation circuits & 3 & 3 & 4 & 25 \\
  Shots per basis & 512 & 1024 & 1024 & 1024 \\

  \textit{SPSA parameters:} & & & & \\
  $a_0$    & 0.1   & 0.1   & calibrated$^\dagger$ & calibrated\\
  $c_0$    & 0.1   & 0.1   & 0.1   & 0.1 \\
  $\alpha$ & 0.602 & 0.602 & 0.602 & 0.602 \\
  $\gamma$ & 0.101 & 0.101 & 0.101 & 0.101 \\
  $A$      & 10    & 20    & 0     & 0 \\

  \hline
  \multicolumn{5}{c}{\textbf{Neural Quantum State}} \\
  \hline

  \textit{Transformer parameters:} & & & & \\
  Layers & 2 & 2 & 2 & 2 \\
  Heads & 4& 4 & 4 & 4 \\
  Internal dimension & 8& 12 & 8 & 8 \\
  Parameter count & 890 & 1766--2006 & 794 & 826 \\

  \hline
  \multicolumn{5}{c}{\textbf{Neural Quantum State Tomography}} \\
  \hline

  Bases & 15 & $2N-1$ & 4 & 11 \\
  Samples per basis & 512 & 512 & 300 & 500 \\
  Batch size & 512 & 512 & 128 & 128 \\
  Learning rate & $10^{-2}$& $10^{-3}$& $10^{-2}$ & $10^{-2}$ \\
  Epochs & 50 & 30 & 100 & 100 \\
   Time (Single CPU) &
  $<1$ minute & $8$ minutes (16 sites) & $30$ seconds & $6$ minutes  \\
  \hline
  \multicolumn{5}{c}{\textbf{Variational Monte Carlo}} \\
  \hline

  Iterations & 400 & 3200 & 1000 & 1200 \\
  Batch size & 512 & $2^{8} - 2^{14}$ & 256 & 256 \\
  Learning rate&
    $10^{-2}$ &
    $3\times10^{-3}$ then decreased by 10 times after 1600 and 2400 iterations &
    $10^{-2}$ &
    $10^{-2}$ \\
  Regularization&
    0.1 for 200 iterations, then 0 &
    $25.6/(2^N)$ decreasing linearly for 1000 iterations, then 0 &
    0.05 for 600 iterations, then 0 &
    0.05 for 600 iterations, then 0
  \\
 Time (Single CPU)  &
  $<1$ minute & $6$ hours (16 sites) & $10$ seconds & $20$ seconds \\
  \hline
\end{tabularx}}

\caption{ \textbf{Hyperparameter values for neural error mitigation components
$|$} Presented are the hyperparameter values of the neural quantum state, VQE
training, neural quantum state tomography, and variational Monte Carlo for each
system studied. $^\dagger$The $a_0$ parameter for each \H2 and LiH variational
circuit is calibrated~\cite{kandala2017hardware} in Qiskit's VQE function.}
\label{tab:hyperparameters}
\end{table}

\clearpage

\end{document}